\documentclass[twocolumn,aps,pra,superscriptaddress]{revtex4-1}
\usepackage{graphicx}
\usepackage{amsmath}
\usepackage{amssymb}
\usepackage{braket}
\usepackage{bm}
\usepackage[colorlinks=true,linkcolor=blue,citecolor=blue,urlcolor=blue]{hyperref}
\usepackage{subfigure}
\usepackage{graphicx}
\usepackage{physics}

\begin{document}

\title{Analytical solution for the spectrum of two ultracold atoms in a
completely anisotropic confinement}

\author{Yue Chen}

\affiliation{Department of Physics, Renmin University of China, Beijing, 100872,
China}

\author{Da-Wu Xiao}

\affiliation{Beijing Computational Science Research Center, Beijing, 100193, China}

\author{Ren Zhang}

\affiliation{School of Science, Xi'an Jiaotong University, Xi'an, 710049, China}

\author{Peng Zhang}
\email{pengzhang@ruc.edu.cn}


\affiliation{Department of Physics, Renmin University of China, Beijing, 100872,
China}

\affiliation{Beijing Computational Science Research Center, Beijing, 100193, China}

\begin{abstract}
We study the system of two ultracold atoms in a three-dimensional (3D)
or two-dimensional (2D) completely anisotropic harmonic trap. We derive the
algebraic equation $J_{\rm 3D}(E)=1/a_{\rm 3D}$ ($J_{\rm 2D}(E)=\ln a_{\rm 2D}$)
for the eigen-energy $E$ of this system in the 3D (2D) case, with
$a_{\rm 3D}$ and $a_{\rm 2D}$ being the corresponding $s$-wave scattering lengths,
and provide the analytical expressions of the functions $J_{\rm 3D}(E)$
and $J_{\rm 2D}(E)$. In previous researches this type of equation
was obtained for spherically or axially symmetric
harmonic traps (T. Busch, {\it et. al.}, Found. Phys. {\bf 28}, 549
(1998); Z. Idziaszek and T. Calarco, Phys. Rev. A {\bf 74}, 022712
(2006)). However, for our cases with a completely anisotropic trap, 
only the equation for the ground-state energy of some cases has been derived (J. Liang and
C. Zhang, Phys. Scr. {\bf 77}, 025302 (2008)). Our results in this work are applicable
for arbitrary eigen-energy of this system, and can be used for the
studies of dynamics and thermal-dynamics of interacting ultracold atoms
in this trap, {\it e.g.}, the calculation of the 2nd virial coefficient
or the evolution of two-body wave functions. In addition, our approach
for the derivation of the above equations can also be used for other
two-body problems of ultracold atoms. 
\end{abstract}
\maketitle

\section{Introduction}

The two-body problems of trapped interacting ultracold atoms are basic
problems in cold atom physics \cite{busch,calarco,calarco2,liang,Blume_review,szenz09,szenz11,szenz13,szenz16,Liyou,Schmelch1,Schmelch2,Schmelch3}.
They are of broad interest because of the
following reasons. First, the two-body systems are ``minimum'' interacting
systems of trapped ultracold atoms, and one can obtain a primary understanding
for the interaction physics of an ultracold gas from the analysis
of such systems \cite{Jochim}.
Second, the solutions to these problems can be
directly used to calculate some important  few- or many-body quantities \cite{blume02,HuiHu_virial,blume08,blume10,blume12,xiaji-virial,blume14,XYYin14,shiguo-virial,blume15,blume18,XYYin20}, 
\textit{e.g.}, the 2nd virial coefficient which determines the high-temperature properties
of the ultracold gases. Third, these systems have been already realized
in many experiments\cite{Jochim,Scazza2014,Cappellini2014,Norcia18,Cooper18,Cappellini2019,guan2b,Ni1,Ni2,Ni3,XDHe,nano1,nano2,nano3}
where the trap of the two atoms can be created via an optical lattice site \cite{Scazza2014,Cappellini2014,Cappellini2019},
an optical tweezer \cite{guan2b,Ni1,Ni2,Ni3,XDHe},
 or nano-structure \cite{nano1,nano2,nano3}.
In these experiments, by measuring or controlling
the energy spectrum or dynamics of these two atoms, one can, \textit{e.g.},
create a cold molecule in
an optical tweezer \cite{Ni1,Ni2},
derive the parameters of inter-atomic interaction potential \cite{Scazza2014,Cappellini2014,Cappellini2019,Ni3}
or study various dynamical effects such as the interaction-induced
density oscillation \cite{guan2b}.
Theoretical results for the corresponding two-body
problems are very important for these experimental studies.

The most fundamental systems of two trapped ultra-cold atoms  are
the ones with a harmonic trap which has the same
frequencies for each atom, and  a $s$-wave short-range
inter-atomic interaction. For these systems, the center-of-mass (CoM)
and relative motion of the two atoms can be decoupled with each other,
and the CoM motion is just the same as a harmonic oscillator. Nevertheless,
the dynamics of the relative motion of these two atoms is nontrivial.
For the three-dimensional (3D) systems, T. Busch, $\textit{et. al.}$
\cite{busch} and Z. Idziaszek and T. Calarco \cite{calarco2,calarco} studied
the cases with a spherically and axially symmetric harmonic trap, respectively.
They show that the eigen-energy $E$ of the relative motion satisfies
an algebraic equation with the form 
\begin{equation}
J_{{\rm 3D}}(E)=\frac{1}{a_{{\rm 3D}}},\label{3de}
\end{equation}
with $a_{{\rm 3D}}$ being the 3D $s$-wave scattering length, and
provided the analytical expressions of the function $J_{{\rm 3D}}(E)$
for these two cases \cite{busch,calarco2,calarco}, as shown in Table I, respectively.
Moreover, for the two-dimensional (2D) systems with an isotropic harmonic
trap, T. Busch, \textit{et. al.} obtained the equation

\global\long\def\arraystretch{2.2}
 
\begin{table*}
\vspace{22pt}
 \centering %
\begin{tabular}{|c|c|c|c|}
\hline 
\hspace{2pt} type \hspace{2pt}  & $E$  & $J_{{\rm 3D}}(E)$  & {Ref.} \tabularnewline
\hline 
\begin{tabular}{@{}c@{}}
\hspace{2pt} $\omega_{x}=\omega_{y}=\omega_{z}$ \hspace{2pt} \tabularnewline
\end{tabular} & %
\begin{tabular}{@{}c@{}}
\hspace{2pt} arbitrary $E$ \hspace{2pt} \tabularnewline
\end{tabular} & $\sqrt{2}\frac{\Gamma\left(-\frac{E}{2}+\frac{3}{4}\right)}{\Gamma\left(-\frac{E}{2}+\frac{1}{4}\right)}$  & Ref.~\cite{busch} \tabularnewline
\hline 
\begin{tabular}{@{}c@{}}
$\omega_{x}=\omega_{y}\neq\omega_{z}$ \tabularnewline
\end{tabular} & %
\begin{tabular}{@{}c@{}}
arbitrary $E$ \tabularnewline
\end{tabular} & %
\begin{tabular}{@{}c@{}}
\hspace{2pt} $-\frac{\eta_{x}}{\sqrt{2}}\sum\limits _{n={0}}^{\infty}\left[\frac{\Gamma\left(-\frac{E-E_{0}}{2}+n\eta_{x}\right)}{\Gamma\left(\frac{1}{2}-\frac{E-E_{0}}{2}+n\eta_{x}\right)}-\frac{1}{\sqrt{\eta_{x}}\sqrt{n+1}}\right]-\sqrt{\frac{\eta_{x}}{2}}\zeta\left(\frac{1}{2}\right)$
\vspace{4pt}
 \tabularnewline
\end{tabular} & Ref.~\cite{calarco2} $(\ast)$ \tabularnewline
\hline 
\begin{tabular}{@{}c@{}}
$\omega_{x}\neq\omega_{y}\neq\omega_{z}$ \tabularnewline
\end{tabular} & %
\begin{tabular}{@{}c@{}}
$E<E_{0}$ \tabularnewline
\end{tabular} & %
\begin{tabular}{@{}c@{}}
$-\frac{1}{2\sqrt{\pi}}\int_{0}^{\infty}dt\left\{ \frac{\sqrt{\eta_{x}\eta_{y}}\exp[{(E-E_{0})t}/{2}]}{\sqrt{1-e^{-\eta_{x}t}}\sqrt{1-e^{-\eta_{y}t}}\sqrt{1-e^{-t}}}-\frac{1}{t^{3/2}}\right\} $
\vspace{4pt}
 \tabularnewline
\end{tabular} & Ref.~\cite{liang} \tabularnewline
\hline 
$\omega_{x}\neq\omega_{y}\neq\omega_{z}$  & %
\begin{tabular}{@{}c@{}}
arbitrary $E$\tabularnewline
\end{tabular} & Eq.~(\ref{dg}) or Eq. (\ref{dg22})  & this work\tabularnewline
\hline 
\end{tabular}\label{bs} \caption{Expressions of the function $J_{{\rm 3D}}(E)$ of Eq. (\ref{3de})
for various 3D traps. Here $\omega_{\alpha}$ ($\alpha=x,y,z$) is
the trapping frequency in the $\alpha$-direction, and we use the natural
unit $\hbar=2\mu=\omega_{z}=1$ with $\mu$ being the reduced mass
of the two atoms. The parameters $\eta_{x}$ and $\eta_{y}$ are defined
as $\eta_{x}=\omega_{x}/\omega_{z}$ and $\eta_{y}=\omega_{y}/\omega_{z}$,
respectively, $E_{0}=(\eta_{x}+\eta_{y}+1)/2$ is the ground-state energy of the relative motion in the non-interacting case, and $\Gamma(z)$ and $\zeta(z)$ are the Gamma and Riemann zeta function, respectively. 
The two expressions (\ref{dg}) and (\ref{dg22}) of $J_{\rm 3D}(E)$ given by this work are mathematically equivalent with each other.
($\ast$): Notice that there are typos in Eqs. (21, 23) of Ref.~\cite{calarco2}.  }
\end{table*}

\begin{equation}
J_{{\rm 2D}}(E)=\ln a_{{\rm 2D}},\label{2de}
\end{equation}
for the relative-motion eigen-energy $E$, where $a_{{\rm 2D}}$ is
the 2D $s$-wave scattering length, and derive expression of the function
$J_{{\rm 2D}}(E)$ \cite{busch}. With these results one
can easily obtain the complete energy spectrum of these two-atom systems,
as well as the corresponding eigen-states. Thus, these results have
been widely used in the researches of cold atom physics \cite{Blume_review,Jochim,Scazza2014,Cappellini2014,Cappellini2019,Ni3,Ospelkaus06,Stoferle06,Mark11,Riegger18,Chapurin19}.

However, for  two atoms trapped in  a 3D completely anisotropic harmonic
confinement,  the algebraic equation for {\it arbitrary} 
eigen-energy  of relative-motion has not been obtained so far. Only the equation of the \textit{ground-state} energy of a system with positive scattering length has been obtained 
 (i.e., the eigen-energy $E$ which satisfies $E<E_0$, with $E_0$ being the relative ground-state energy for the non-interacting case) in Ref. \cite{liang}, by J. Liang and C. Zhang in 2008.
 This equation  has the form of Eq.~(\ref{3de}), and the corresponding function $J_{\rm 3D}(E)$ is also
 shown in Table I. On the other hand, since this kind of confinements
are easy to be experimentally prepared, and are thus used in many
experiments of ultracold gases \cite{Scazza2014,Cappellini2014,Norcia18,Cooper18,Cappellini2019,Ni3,Chapurin19}, it would be helpful
if we can derive a general equation satisfied by all the eigen-energies.

In this work we  derive the equation for arbitrary
eigen-energy of two atoms in a completely anisotropic harmonic trap.
We show that this equation also has the form of Eq. ($\ref{3de}$)
and Eq. (\ref{2de}) for the 3D and 2D cases, respectively, and provide
the corresponding analytical expressions of the functions $J_{{\rm 3D}}(E)$ (Table I)
and $J_{{\rm 2D}}(E)$ (Eq. (\ref{j2de}) or Eq. (\ref{dg2d22})). As the aforementioned results \cite{busch,calarco2,calarco}
for the spherically and axially symmetric traps, our equations are
useful for the studies of various thermal-dynamcial or dynamical properties
of ultracold gases in the completely anisotropic harmonic confinements.
Furthermore, our calculation approach used in this work can also be
generalized to other two-body problems of ultracold atoms in complicated
confinements.

The remainder of this paper is organized as follows. In Sec. II and
Sec. III, we derive the equations for eigen-energies of atoms in 3D
and 2D completely anisotropic harmonic traps, respectively.
In Sec. IV we discuss how to generalize our approach to other problems. 
 A brief summary
and some discussions are given in Sec. V. Some details of our calculations
are shown in the appendix.

\section{3D systems}

We consider two ultracold atoms $1$ and $2$ in a 3D completely anisotropic
harmonic trap which has the same frequencies for each atom. Here we
denote $\omega_{\alpha}$ ($\alpha=x,y,z$) as the trapping frequency
in the $\alpha$-direction, which satisfy 
\begin{equation}
\omega_{x}\neq\omega_{y}\neq\omega_{z}.\label{ome}
\end{equation}
For convenience, in this work we use the natural unit 
\begin{equation}
\hbar=2\mu=\omega_{z}=1,\label{nu}
\end{equation}
with $\mu$ being the reduced mass of the two atoms. We further introduce
the aspect ratios
\begin{eqnarray}
\eta_{x}=\frac{\omega_{x}}{\omega_{z}};\ \eta_{y}=\frac{\omega_{y}}{\omega_{z}};\ \eta_{z}=\frac{\omega_{z}}{\omega_{z}}=1,
\end{eqnarray}
where $\eta_{x}$ and $\eta_{y}$ describe the anisotropy of the trapping
potential.

As mentioned in Sec. I, for this system we can separate out the CoM
degree of freedom of these two atoms and focus on the inter-atomic
relative motion. The Hamiltonian operator of our problem is given
by 
\begin{equation}
\hat{H}=\hat{H}_{0}+\hat{V}_{I}.\label{hrel}
\end{equation}
Here $\hat{H}_{0}$ is the free Hamiltonian for the relative motion
and can be expressed as 
\begin{equation}
\hat{H}_{0}=\hat{{\bf p}}^{2}+\frac{1}{4}\left[\eta_{x}^{2}\hat{x}^{2}+\eta_{y}^{2}\hat{y}^{2}+\hat{z}^{2}\right],\label{h03d}
\end{equation}
with $\hat{{\bf p}}$ and $\hat{{\bf r}}\equiv(\hat{x},\hat{y},\hat{z})$
being the relative momentum and coordinate operators, respectively.
In Eq. (\ref{hrel}) $\hat{V}_{I}$ is the inter-atomic interaction
operator, which is modeled as the $s$-wave Huang-Yang pseudo potential.
Explicitly, for any state $|\psi\rangle$ of the realtive motion we
have 
\begin{equation}
\langle{\bf r}|\hat{V}_{I}|\psi\rangle=4\pi a_{{\rm 3D}}\delta({\bf r})\frac{\partial}{\partial r}\left[r\cdot\langle{\bf r}|\psi\rangle\right],\label{hy}
\end{equation}
with $|{\bf r}\rangle$ being the eigen-state of the relative-coordinate
operator $\hat{{\bf r}}$ with eigen-value ${\bf r}$, $r=|{\bf r}|$,
and $a_{{\rm 3D}}$ being the 3D $s$-wave scattering length. 

For our system the parity with respect  to the spatial inversion ${\bf r}\rightarrow-{\bf r}$ is conserved, and the contact pseudo potential $\hat{V}_{I}$ only operates on the states with even parity. 
Therefore, in this work we only consider the eigen-energies and eigen-states of  $\hat{H}$ in the even-parity subspace.

Now we deduce the algebraic equation for the eigen-energy $E$ of
the total Hamiltonian $\hat{H}$. 
We begin from the schr$\ddot {\rm o}$edinger equation 
\begin{equation}
\left[{\hat{H}}_{0}+\hat{V}_{I}\right]|\Psi\rangle=E|\Psi\rangle,\label{eigen}
\end{equation}
satisfied 
by $E$ and the corresponding
eigen-state $|\Psi\rangle$ of $\hat{H}$. This equation can be re-expressed as 
\begin{equation}
|\Psi\rangle=\frac{1}{E-\hat{H}_{0}}\hat{V}_{I}|\Psi\rangle.\label{e2}
\end{equation}
Using Eq. (\ref{hy}) we find that Eq. (\ref{e2}) yields 
\begin{equation}
\langle{\bf r}|\Psi\rangle=4\pi a_{{\rm 3D}}G_{0}(E,{\bf r})\left[\left.\frac{\partial}{\partial r}\left[r\cdot\langle{\bf r}|\Psi\rangle\right]\right|_{{\bf r}={\bf 0}}\right],\label{eq2}
\end{equation}
where $G_{0}(E,{\bf r})$ is the Green's function of the free Hamiltonian
${\hat{H}}_{0}$, which is defined as 
\begin{equation}
G_{0}(E,{\bf r})=\langle{\bf r}|\frac{1}{E-\hat{H}_{0}}|{\bf 0}\rangle.\label{green0}
\end{equation}

We can derive the the equation for the eigen-energy $E$ by doing
the the operation $\left.\frac{\partial}{\partial r}(r\cdot)\right|_{{\bf r}=0}$
on both sides of Eq. (\ref{eq2}). In this operation, without loss
of generality, we choose ${\bf r}=z{\bf e}_{z}$ and ${\bf e}_{z}$
being is the unit vector along the $z$-direction. Then we find that
$E$ satisfies 
\begin{eqnarray}
J_{{\rm 3D}}(E)=\frac{1}{a_{{\rm 3D}}},\label{11}
\end{eqnarray}
which is just Eq. (\ref{3de}) of Sec. I, with the function $J_{{\rm 3D}}(E)$
being defined as 
\begin{equation}
J_{{\rm 3D}}(E)=4\pi\left\{ \left.\frac{\partial}{\partial|z|}\left[|z|\cdot G_{0}(E,z{\bf e}_{z})\right]\right|_{z\rightarrow0}\right\} .\label{fe}
\end{equation}

\bigskip{}

\subsection{Expression of $J_{{\rm 3D}}(E)$}

Next we derive the expression of the function $J_{{\rm 3D}}(E)$.
For our system, the ground-sate energy of the relative motion of two non-interacting atoms is 
\begin{eqnarray}
E_{0}=\frac{1}{2}(\eta_{x}+\eta_{y}+1).
\end{eqnarray}
In the following we first consider the case with $E<E_{0}$, which
was also studied by Ref.~\cite{liang}, and then investigate the
general case with arbitrary $E$.

\subsubsection{Special Case: $E<E_{0}$}

When $E<E_{0}$, the Green's function $G_{0}(E,z{\bf e}_{z})$ ($z>0$)
can be expressed as the Laplace transform of the imaginary-time propagator \cite{Yvan,Ren18,Ren19,Xiao2019,Ren20},
i.e., 
\begin{eqnarray}
G_{0}(E,z{\bf e}_{z})=-\int_{0}^{+\infty}K(z,E,\beta)d\beta,\label{ggg2}
\end{eqnarray}
with the function $K(z,E,\beta)$ being defined as 
\begin{eqnarray}
 &  & K(z,E,\beta)\nonumber \\
 & = & e^{\beta E}\langle z{\bf e}_{z}|e^{-\beta\hat{H}_{0}}|{\bf 0}\rangle\nonumber \\
 & = & \exp\left[\beta E-\frac{z^{2}}{4}{\coth\beta}\right]\prod_{\alpha=x,y,z}\sqrt{\frac{\eta_{\alpha}}{4\pi\sinh\left(\eta_{\alpha}\beta\right)}}.\nonumber\\
 \label{kr}
\end{eqnarray}
Here we emphasize  that, when $E<E_{0}$ the function $K(z,E,\beta)$ exponentially
decays to zero in the limit $\beta\rightarrow\infty$, and thus the
integration in Eq. (\ref{ggg2}) converges for any fixed non-zero
$z$. Nevertheless, this integration diverges in the limit $z\rightarrow0$.
That is due to the behavior of the leading term $e^{-z^{2}/(4\beta)}/(4\pi\beta)^{\frac{3}{2}}$
of the function $K(z,E,\beta)$ in the limit $\beta\rightarrow0^{+}$ \cite{Yvan,Ren18,Ren19,Xiao2019,Ren20}.
We can separate this divergence by re-expressing the integration as
\begin{eqnarray}
G_{0}(E,z{\bf e}_{z}) & = & -\int_{0}^{+\infty}d\beta\frac{e^{-\frac{z^{2}}{4\beta}}}{(4\pi\beta)^{\frac{3}{2}}}-\int_{0}^{+\infty}d\beta\tilde{K}(r,E,\beta)\nonumber \\
 & = & -\frac{1}{4\pi|z|}-\int_{0}^{+\infty}d\beta\tilde{K}(z,E,\beta),\label{int3a}
\end{eqnarray}
where 
\begin{equation}
\tilde{K}(z,E,\beta)=K(z,E,\beta)-\frac{e^{-\frac{z^{2}}{4\beta}}}{(4\pi\beta)^{\frac{3}{2}}}.\label{kp}
\end{equation}
In Eq. (\ref{int3a}) the integration $\int_{0}^{+\infty}d\beta\tilde{K}(z,E,\beta)$
uniformly converges in the limit $z\rightarrow0$. Using this result,
we obtain the expansion of $G_{0}(E,z{\bf e}_{z})$ in this limit:
\begin{eqnarray}
\lim_{z\rightarrow0}G_{0}(E,z{\bf e}_{z}) & = & -\frac{1}{4\pi|z|}-\int_{0}^{+\infty}d\beta\tilde{K}(0;E,\beta)+{\cal O}(z).\nonumber \\
\label{gexpa}
\end{eqnarray}
Substituting this result into Eq. (\ref{fe}) and using Eqs. (\ref{kr},
\ref{kp}), we finally obtain the expression 
\begin{eqnarray}
 &  & J_{{\rm 3D}}(E)\nonumber \\
 & = & -\int_{0}^{+\infty}d\beta\left\{ \frac {e^{\beta E}}{2\sqrt{\pi}}\!\!\prod_{\alpha=x,y,z}\sqrt{\frac{\eta_{\alpha}}{\sinh\left(\eta_{\alpha}\beta\right)}}-{1\over 2\sqrt{\pi} }\frac{1}{\beta^{\frac{2}{3}}}\right\} \nonumber \\
 &  & ({\rm for}\ \ E<E_{0}),\label{dga}
\end{eqnarray}
where the the integration converges for $E<E_{0}$, as the one in
Eq.~(\ref{ggg2}). This result was also derived by J. Liang and C.
Zhang in Ref.~\cite{liang}.

\subsubsection{General Case: Arbitrary $E$}

In the general case with arbitrary energy $E$ we cannot directly
use the above result in Eq. (\ref{dga}), because the integration
in this equation diverges for $E>E_{0}$. For the systems with spherically
or axially symmetric confinements, the authors of Refs. \cite{busch,calarco}
successfully found the analytical continuation of this integration
for all real $E$. However, for the current system with completely
anisotropic traps, to our knowledge, so far such analytical continuation
has not been found.

Now we introduce our approach to solve this problem. For convenience, we first define the eigen-energy of
the free Hamiltonian $\hat{H}_{0}$, which is just a Hamiltonian of
a 3D harmonic oscillator, as $E_{{\bf n}}$. Here 
\begin{eqnarray}
{\bf n} & = & (n_{x},n_{y},n_{z}),
\end{eqnarray}
with $n_{\alpha}=0,1,2,...$ ($\alpha=x,y,z$) being the quantum number
of the $\alpha$-direction. It is clear that we have 
\begin{eqnarray}
E_{{\bf n}}\equiv\epsilon_{n_{x}}+\epsilon_{n_{y}}+\epsilon_{n_{z}},
\end{eqnarray}
with 
\begin{eqnarray}
\epsilon_{n_{\alpha}} & = & \left(\frac{1}{2}+n_{\alpha}\right)\eta_{\alpha},\ \ (\alpha=x,y,z).\label{eps}
\end{eqnarray}
In addition, we further denote the eigen-state of $\hat{H}_{0}$
corresponding to $E_{{\bf n}}$ as $|{\bf n}\rangle$.

Similar to above, the key step of our approach is to calculate free
Green's function $G_{0}(E,z{\bf e}_{z})$ defined in Eq.~(\ref{green0}).
Since the result in Eq.~(\ref{ggg2}) cannot be used for our
general case because the integration in this equation diverges for
$E>E_{0}$, we need to find another expression for $G_{0}(E,z{\bf e}_{z})$,
which converges for any $E$. To this end, we separate all the eigen-states
$\{|{\bf n}\rangle\}$ of $\hat{H}_{0}$ into two groups, i.e., the
ones with ${\bf n}\in L_{E}$ and ${\bf n}\in U_{E}$, respectively,
with the sets $L_{E}$ and $U_{E}$ being defined as 
\begin{widetext}
\begin{equation}
L_{E}:\left\{ \left.(n_{x},n_{y},n_{z})\right|n_{x,y,z}=0,1,2...,\ \epsilon_{n_{x}}+\epsilon_{n_{y}}+\frac{1}{2}\leq E\right\} ,\label{ne}
\end{equation}
and 
\begin{equation}
U_{E}:\left\{ \left.(n_{x},n_{y},n_{z})\right|n_{x,y,z}=0,1,2...,\ \epsilon_{n_{x}}+\epsilon_{n_{y}}+\frac{1}{2}>E\right\} ,\label{ngeq}
\end{equation}
respectively. It is clear that we have 
\begin{equation}
E_{{\bf n}}>E,\ \ \ {\rm for\ all\ states\ with}\ {\bf n}\in U_{E}.\label{ee2a}
\end{equation}
Furthermore, we can re-express the free Green's operator $1/[E-\hat{H}_{0}]$
as 
\begin{eqnarray}
\frac{1}{E-\hat{H}_{0}} & = & \sum_{{\bf n}\in U_{E}}\frac{|{\bf n}\rangle\langle{\bf n}|}{E-E_{{\bf n}}}+\sum_{{\bf n}\in L_{E}}\frac{|{\bf n}\rangle\langle{\bf n}|}{E-E_{{\bf n}}}\label{green}\label{sep}\\
 & = & -\int_{0}^{+\infty}d\beta\left[\sum_{{\bf n}\in U_{E}}|{\bf n}\rangle\langle{\bf n}|e^{-\beta\left(E_{{\bf n}}-E\right)}\right]+\sum_{{\bf n}\in L_{E}}\frac{|{\bf n}\rangle\langle{\bf n}|}{E-E_{{\bf n}}}.\label{g2}
\end{eqnarray}
Due to the fact (\ref{ee2a}), the integration in 
of Eq. (\ref{g2}) \textit{converges}. Moreover, the integrand
in Eq. (\ref{g2}) can be re-written as 
\begin{eqnarray}
\sum_{{\bf n}\in U_{E}}|{\bf n}\rangle\langle{\bf n}|e^{-\beta\left(E_{{\bf n}}-E\right)} & = & \sum_{{\bf n}}|{\bf n}\rangle\langle{\bf n}|e^{-\beta\left(E_{{\bf n}}-E\right)}-\sum_{{\bf n}\in L_{E}}|{\bf n}\rangle\langle{\bf n}|e^{-\beta\left(E_{{\bf n}}-E\right)}\nonumber \\
 & = & e^{\beta E}e^{-\beta\hat{H}_{0}^{{\rm (3D)}}}-\sum_{{\bf n}\in L_{E}}|{\bf n}\rangle\langle{\bf n}|e^{-\beta\left(E_{{\bf n}}-E\right)},\label{ss2}
\end{eqnarray}
where we have used $e^{-\beta\hat{H}_{0}^{{\rm (3D)}}}=\sum_{{\bf n}}|{\bf n}\rangle\langle{\bf n}|e^{-\beta E_{{\bf n}}}$.
Substituting Eq. (\ref{ss2}) into Eq. (\ref{g2}) and then into Eq.
(\ref{green0}), we obtain 
\begin{equation}
G_{0}(E,z{\bf e}_{z})=-\int_{0}^{+\infty}d\beta\left[K(z;E,\beta)-F(z;E,\beta)\right]+Q(z;E),\label{g3}
\end{equation}
where $K(z;E,\beta)$ is defined in Eq. (\ref{kr}), and the functions
$F(z;E,\beta)$ and $Q(z;E,\beta)$ are defined as 
\begin{eqnarray}
F(z;E,\beta) & = & \sum_{{\bf n}\in L_{E}}\langle{z{\bf e}}_{z}|{\bf n}\rangle\langle{\bf n}|{\bf 0}\rangle e^{-\beta\left(E_{{\bf n}}-E\right)};\label{fr}\\
Q(z;E,\beta) & = & \sum_{{\bf n}\in L_{E}}\frac{\langle{z{\bf e}}_{z}|{\bf n}\rangle\langle{\bf n}|{\bf 0}\rangle}{E-E_{{\bf n}}}.\label{qr}
\end{eqnarray}
Due to the convergence of the integration in Eq. (\ref{g2}), the
integration in Eq. (\ref{g3}) also converges for non-zero $z$, no
matter if $E>E_{0}$ or $E<E_{0}$. Therefore, Eq. (\ref{g3}) is
the convergent expression of $G_{0}(E,z{\bf e}_{z})$ for arbitrary
$E$.

Furthermore, similar to in Sec. II. A, the integration in Eq. (\ref{g3})
diverges in the limit $|z|\rightarrow0$, due to the leading term
$e^{-\frac{z^{2}}{4\beta}}/(4\pi\beta)^{\frac{3}{2}}$ of the integrand in the limit $\beta\rightarrow0^{+}$. We can separate this
divergence via the technique used in Eqs. (\ref{int3a}-\ref{gexpa}),
i.e., re-express this integration as
\begin{eqnarray}
-\int_{0}^{+\infty}d\beta\left[K(z;E,\beta)-F(z;E,\beta)\right] & = & -\int_{0}^{+\infty}d\beta\frac{e^{-\frac{z^{2}}{4\beta}}}{\left(4\pi\beta\right)^{\frac{3}{2}}}-\int_{0}^{+\infty}d\beta[\tilde{K}(z;E,\beta)-F(z;E,\beta)]\nonumber\\
 & = & -\frac{1}{4\pi|z|}-\int_{0}^{+\infty}d\beta[\tilde{K}(z;E,\beta)-F(z;E,\beta)],
\end{eqnarray}
with the function $\tilde{K}(z;E,\beta)$ being defined in Eq. (\ref{kp}).
Using this technique we obtain 
\begin{equation}
\lim_{|z|\rightarrow0}G^{(0)}(E,z{\bf e}_{z})=-\frac{1}{4\pi|z|}+\left[W_{{\rm 3D}}(E)+\int_{0}^{+\infty}I_{{\rm 3D}}(E,\beta)d\beta\right]+{\cal O}(z),\label{gexp}
\end{equation}
where the $z$-independent functions $W_{{\rm 3D}}(E)$ and $I_{{\rm 3D}}(E,\beta)$
are defined as $W_{{\rm 3D}}(E)\equiv Q(0,E,\beta)$ and $I_{{\rm 3D}}(E,\beta)\equiv-\tilde{K}(0;E,\beta)+F(0;E,\beta)$,
respectively. The straightforward calculations (Appendix A) show that
they can be expressed as 
\begin{equation}
W_{{\rm 3D}}(E)=-\frac{\pi}{2}\sqrt{\frac{\eta_{x}\eta_{y}}{2}}\sum_{(n_{x},n_{y})\in C_{E}^{{\rm (3D)}}}\left\{ \frac{2^{n_{x}+n_{y}-1}\Gamma\left[\frac{1}{4}-\frac{\left(E-\epsilon_{n_{x}}-\epsilon_{n_{y}}\right)}{2}\right]}{\Gamma\left(\frac{1-n_{x}}{2}\right)^{2}\Gamma\left(\frac{1-n_{y}}{2}\right)^{2}\Gamma(n_{x}+1)\Gamma(n_{y}+1)\Gamma\left[\frac{3}{4}-\frac{\left(E-\epsilon_{n_{x}}-\epsilon_{n_{y}}\right)}{2}\right]}\right\} ,\label{wee}
\end{equation}
and 
\begin{eqnarray}
I_{{\rm 3D}}(E,\beta) & = & -e^{\beta E}\prod_{\alpha=x,y,z}\sqrt{\frac{\eta_{\alpha}}{4\pi\sinh\left(\eta_{\alpha}\beta\right)}}+\left(\frac{1}{4\pi\beta}\right)^{\frac{3}{2}}\nonumber \\
 &  & +\sqrt{\frac{\pi\eta_{x}\eta_{y}}{8\sinh\beta}}\sum_{(n_{x},n_{y})\in C_{E}^{{\rm (3D)}}}\left\{ \frac{2^{n_{x}+n_{y}-\frac{1}{2}}}{\Gamma\left(\frac{1-n_{x}}{2}\right)^{2}\Gamma\left(\frac{1-n_{y}}{2}\right)^{2}\Gamma(n_{x}+1)\Gamma(n_{y}+1)}e^{\beta(E-\epsilon_{n_{x}}-\epsilon_{n_{y}})}\right\} ,\label{iee}
\end{eqnarray}
where $\Gamma(z)$ is the Gamma function and $C_{E}^{{\rm (3D)}}$ is a set of \textit{two}-dimensional number
array $(n_{x},n_{y})$, which is defined as 
\begin{equation}
C_{E}^{{\rm (3D)}}:\left\{ \left.(n_{x},n_{y})\right|n_{x,y}=0,2,4,6,...;\ \epsilon_{n_{x}}+\epsilon_{n_{y}}+\frac{1}{2}\leq E\right\} .\label{lep}
\end{equation}
Finally, substituting Eq. (\ref{gexp}) into Eq. (\ref{fe}), we obtain the expression of $J_{{\rm 3D}}(E)$.
\begin{eqnarray}
J_{{\rm 3D}}(E) & = & 4\pi\left[W_{{\rm 3D}}(E)+\int_{0}^{+\infty}I_{{\rm 3D}}(E,\beta)d\beta\right],\label{dg}
\end{eqnarray}
with $W_{{\rm 3D}}(E)$ and $I_{{\rm 3D}}(E,\beta)$ being given by Eqs. (\ref{wee}) and  (\ref{iee}), respectively. 
Notice that  
the summations in Eqs. (\ref{wee}, \ref{iee}) are done for finite terms, and the integration in Eq. (\ref{dg}) converges. Therefore, using Eqs.~(\ref{dg}, \ref{wee}, \ref{iee}) one can efficiently  calculate $J_{{\rm 3D}}(E)$. 


It is
clear that Eq. (\ref{11}) for the eigen-energy and 
the expression (\ref{dg}) of the function $J_{\rm 3D}(E)$
are correct for not only the systems in completely anisotropic
traps but also the ones in  spherically or axially symmetric traps. For the latter two cases  Eq. (\ref{dg}) is equivalent to the results derived by Ref. \cite{busch, calarco}, which are shown in Table I.

\subsection{Techniques for Fast Calculation of $J_{{\rm 3D}}(E)$}

There are some techniques which may speed up the calculation for the function $J_{{\rm 3D}}(E)$. First, as shown in Appendix B, $J_{{\rm 3D}}(E)$ can be re-expressed as
\begin{eqnarray}
J_{{\rm 3D}}(E) & = & 4\pi\left[\int_{0}^{\Lambda}A_{{\rm 3D}}(E,\beta)d\beta+B^{(1)}_{\rm 3D}(E, \Lambda)
 +B^{(2)}_{\rm 3D}(E,\Lambda) +\qty(\frac{1}{2\pi})^{3/2}\frac{1}{\sqrt{2\Lambda}}\right],
\label{dg22}
\end{eqnarray}
where $\Lambda$ is an arbitrary finite positive number, and the functions $A_{{\rm 3D}}(E,\beta)$ and $B^{(1,2)}_{{\rm 3D}}(E,\Lambda)$ are defined as
\begin{eqnarray}
A_{{\rm 3D}}(E,\beta)&=& -e^{\beta E}\prod_{\alpha=x,y,z}\sqrt{\frac{\eta_{\alpha}}{4\pi\sinh\left(\eta_{\alpha}\beta\right)}}+\left(\frac{1}{4\pi\beta}\right)^{\frac{3}{2}};\label{a3d}\\
&&\nonumber\\
&&\nonumber\\
B^{(1)}_{{\rm 3D}}(E,\Lambda)&=&(-1)\times\!\!\!\!\sum_{(n_{x},n_{y})\in C_{E}^{{\rm (3D)}}}
\left\{\frac{2^{n_{x}+n_{y}-\frac{5}{2}}\sqrt{\pi\eta_{x}\eta_{y}}\quad \Gamma\left({1\over4}-{{E-\epsilon_{n_{x}}-\epsilon_{n_{y}}}\over2}\right)e^{(E-\epsilon_{n_x}-\epsilon_{n_y}-{3\over2})\Lambda}}{\Gamma\left(\frac{1-n_{x}}{2}\right)^{2}\Gamma\left(\frac{1-n_{y}}{2}\right)^{2}\Gamma(n_{x}+1)\Gamma(n_{y}+1)\Gamma\left({5\over4}-{{E-\epsilon_{n_{x}}-\epsilon_{n_{y}}}\over2}\right)}\times\right.
\nonumber\\
&&\nonumber\\
&&\ \ \ \ \ \ \ \ \ \ \ \ \ \ \ \  \ \ \ \ \ \ \left.\sqrt{e^{2\Lambda}-1}\times _{2}\!F^{1}\qty[1,\frac{3}{4}-\frac{E-\epsilon_{n_x}-\epsilon_{n_y}}{2},\frac{5}{4}-\frac{E-\epsilon_{n_x}-\epsilon_{n_y}}{2},
e^{-2\Lambda}]
\begin{array}{c}
\vspace{18pt}
\end{array}
\right\};
\end{eqnarray}
and
\begin{eqnarray}
\ \ \ \ \ \ \ \ \ \ \ B^{(2)}_{{\rm 3D}}(E,\Lambda)&=&\sqrt{\pi\eta_x\eta_y\csch{\Lambda}}\times\nonumber\\
&&\sum_{(n_{x},n_{y})\notin C_{E}^{{\rm (3D)}}}
\frac{2^{n_{x}+n_{y}-1}e^{(E-\epsilon_{n_x}-\epsilon_{n_y}-2)\Lambda}(e^{2\Lambda}-1)\times _{2}\!F^{1}\qty[1,\frac{3}{4}-\frac{E-\epsilon_{n_x}-\epsilon_{n_y}}{2},\frac{5}{4}-\frac{E-\epsilon_{n_x}-\epsilon_{n_y}}{2},
e^{-2\Lambda}]}{\Gamma\left(\frac{1-n_{x}}{2}\right)^{2}\Gamma\left(\frac{1-n_{y}}{2}\right)^{2}\Gamma(n_{x}+1)\Gamma(n_{y}+1)\left[2\qty(E-\epsilon_{n_x}-\epsilon_{n_y})-1\right]},
 \nonumber\\
 \label{b23d}
\end{eqnarray}
respectively, with $ _{2}F^{1}$ being the Hypergeomtric function.
In Appendix B we prove that  (\ref{dg22}) is exactly equivalent to Eqs. (\ref{dg}) for any finite positive $\Lambda$.
Thus, in the numerical calculation for a specific problem one can choose the value of $\Lambda$ by convenience.
\end{widetext}

Due to the following two facts, the calculation of $J_{{\rm 3D}}(E)$ based Eq. (\ref{dg22})  may be faster than the one based Eq. (\ref{dg}):

 {\bf (A)}
 An important difference between the expressions (\ref{dg}) and (\ref{dg22})  is that,
 the integrand $I_{\rm 3D}(E,\beta)$ of Eq. (\ref{dg}) includes a summation $\sum_{(n_x,n_y)\in C_{E}^{\rm (3D)}}$ while the integrand $A_{\rm 3D}(E,\beta)$ of Eq. (\ref{dg22}) does not include any summation. On the other hand, in the numerical calculations for these integrations for a fixed $E$, one needs to calculate the values of the integrands for many points in the $\beta$-axis. Thus, to calculate the term $\int_0^{+\infty} I_{\rm 3D}(E,\beta)d\beta$ of Eq. (\ref{dg}) one need to do the
the summation $\sum_{(n_x,n_y)\in C_{E}^{\rm (3D)}}$ many times, while to calculate the term $\int_0^{\Lambda} A_{\rm 3D}(E,\beta)d\beta$ of Eq. (\ref{dg22}) one does not require to do that.
This advantage is more significant for  the large-$E$ cases where this summation includes a lot of terms.

 {\bf (B)} Furthermore, the terms in the summation $\sum_{(n_{x},n_{y})\notin C_{E}^{{\rm (3D)}}}
$ of Eq. (\ref{b23d}) decays to zero
 in the limits $n_{x,y}\rightarrow\infty$
 and the decaying is  {\it faster than exponential} (Appendix B). Therefore, although this summation
 includes infinite terms, it can converges fast.

Another technique which may be helpful for the fast calculation of $J_{{\rm 3D}}(E)$ is based on direct conclusions of Eq. (\ref{dg}) or Eq. (\ref{dg22}), i.e., for two energies $E_1$ and $E_2$ we have
\begin{eqnarray}
&&J_{{\rm 3D}}(E_2)- J_{{\rm 3D}}(E_1)\nonumber\\
& = &4\pi\left[  \delta W_{{\rm 3D}}+\int_{0}^{+\infty}\delta I_{{\rm 3D}}(\beta)d\beta\right]\label{dg2a}\\
&=&4\pi\left[\int_{0}^{\Lambda}\delta A_{{\rm 3D}}(\beta)d\beta+\delta B^{(1)}_{\rm 3D}(\Lambda)
 +\delta B^{(2)}_{\rm 3D}(\Lambda)\right]
\label{dg22a}
\end{eqnarray}
where $\delta W_{{\rm 3D}}=W_{{\rm 3D}}(E_2)-W_{{\rm 3D}}(E_1)$, $\delta I_{{\rm 3D}}(\beta)=I_{{\rm 3D}}(E_2,\beta)-I_{{\rm 3D}}(E_1,\beta)$, $\delta A_{\rm 3D}(\beta)=A_{\rm 3D}(E_2,\beta)-A_{\rm 3D}(E_1,\beta)$ and $\delta B^{(1,2)}_{\rm 3D}(\Lambda)=B^{(1,2)}_{\rm 3D}(E_2,\Lambda)- B^{(1,2)}_{\rm 3D}(E_1,\Lambda)$.
Therefore, if the value of $J_{{\rm 3D}}(E_1)$ is already derived, one can calculate $J_{{\rm 3D}}(E_2)$ by either Eq.~(\ref{dg}) (Eq. (\ref{dg22}))
or Eq.~(\ref{dg2a}) (Eq. (\ref{dg22a})), and it is possible that
 the summations and  integration in Eq.~(\ref{dg2a}) (Eq. (\ref{dg22a})) can converge faster.

\begin{figure*}
\centering
\begin{minipage}{6cm}
\includegraphics[width=6cm]{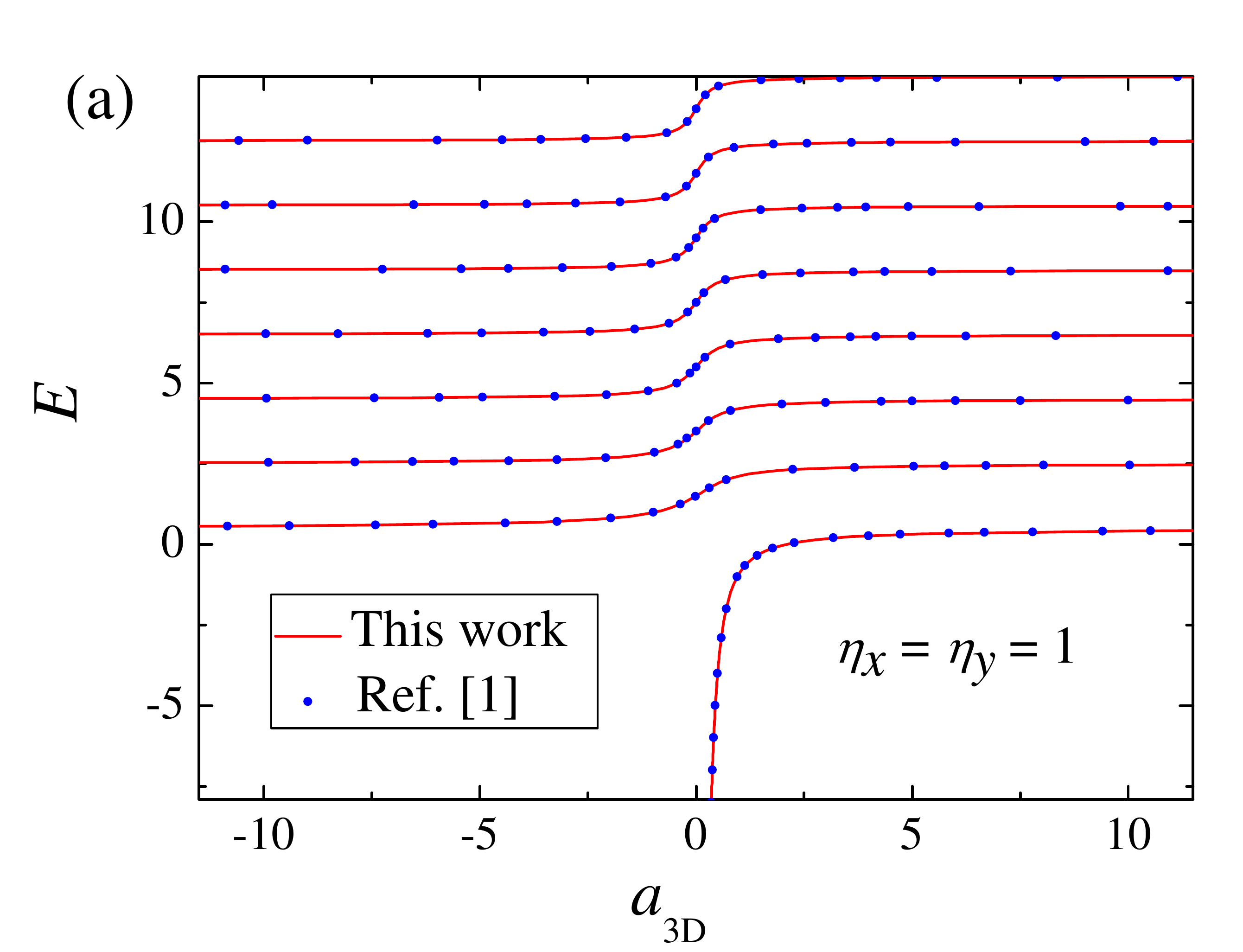} 
\end{minipage}
\begin{minipage}{6cm}
\includegraphics[width=6cm]{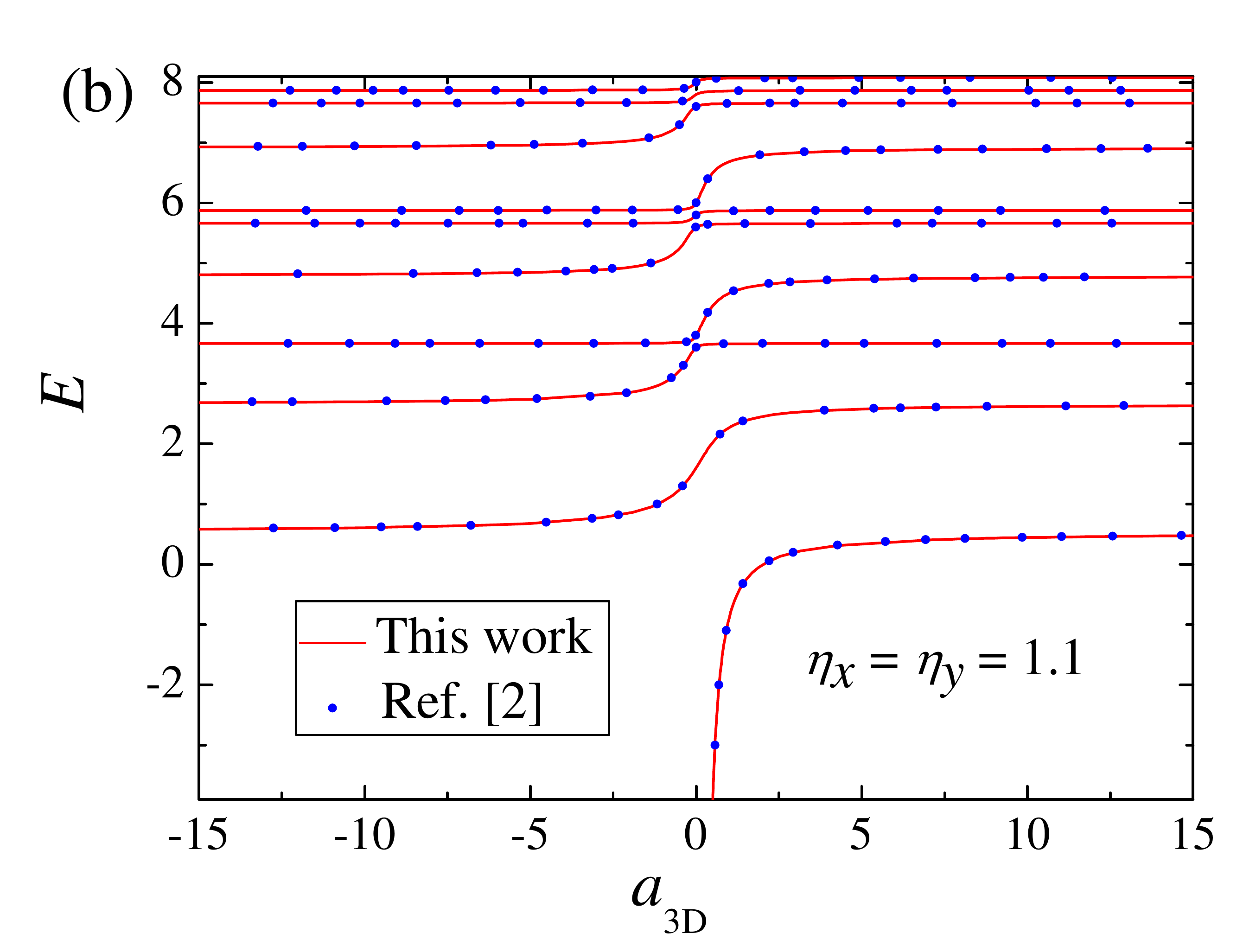} 
\end{minipage}
\begin{minipage}{6cm}
\includegraphics[width=6cm]{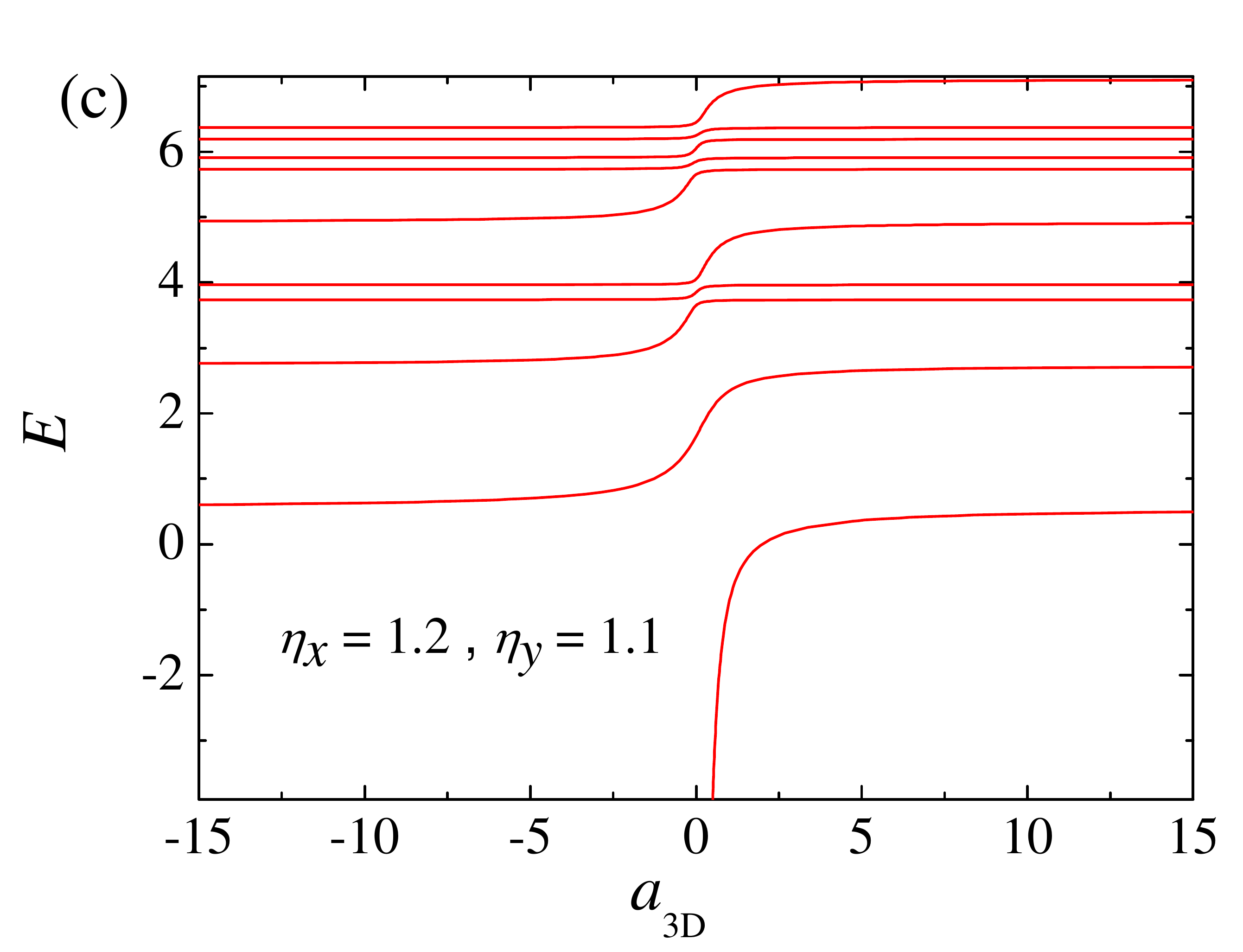} 
\end{minipage}
\begin{minipage}{6cm}
\includegraphics[width=6cm]{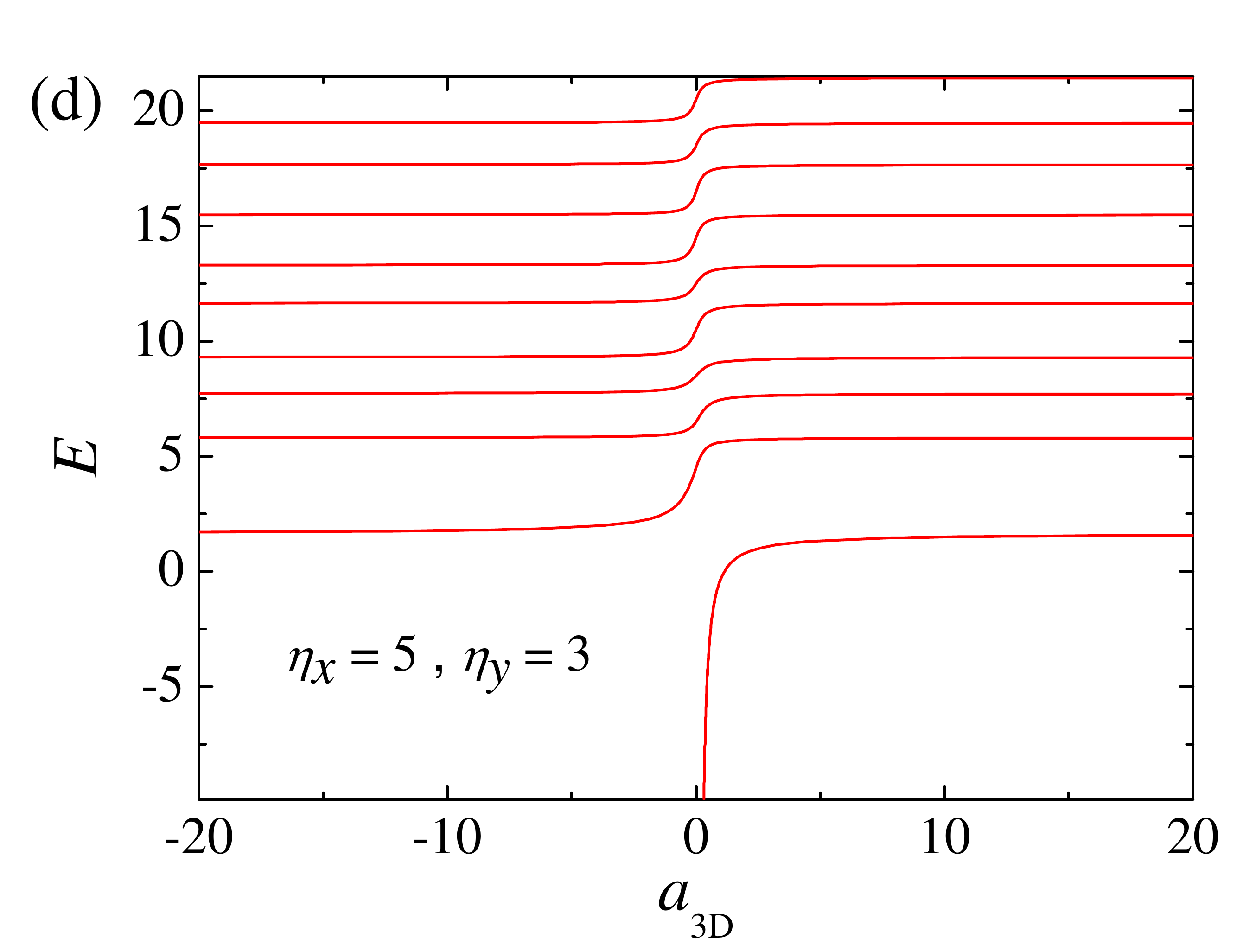} 
\end{minipage}
\caption{(color online)  The energy spectrum of the relative motion of two atoms in a 3D harmonic trap with aspect rations $(\eta_x=\eta_y=1)$ (a),  $(\eta_x=\eta_y=1.1)$ (b), $(\eta_x=1.2, \eta_y=1.1)$ (c)  and $(\eta_x=5, \eta_y=3)$ (d). We show the results given by Eq. (\ref{3de}) and the expression (\ref{dg}) of function $J_{\rm 3D}(E)$ with solid lines. In (a) and (b) we also show the results given by Refs. \cite{busch} and \cite{calarco} with blue dots, respectively. Here we use the natural unit $\hbar=2\mu=\omega_z=1$.}
\label{3d}
\end{figure*}

\subsection{Energy Spectrum}


By solving Eq. (\ref{11}) with the expression (\ref{dg}) of $J_{\rm 3D}(E)$,
we can derive the complete eigen-energy
spectrum of the complete Hamiltonian $\hat{H}$ of the
two-atom relative motion. 
In Fig.~\ref{3d}(a-d) we show the energy spectrum of the relative
motion of two atoms in 3D harmonic traps with various aspect ratios $\eta_{x,y}$, which is derived via Eqs. (\ref{11}, \ref{dg}). In Fig.~\ref{3d}(a) and
(b) we consider the traps with spherical symmetry ($\eta_{x}=\eta_{y}=1$)
and axial symmetry ($\eta_{x}=\eta_{y}=1.1$), respectively, and show that
the results given by our approach are same as the ones from the methods in Refs. \cite{busch}
and \cite{calarco}. In Fig.~\ref{3d}(c) and (d) we show the results
for 3D completely anisotropic traps whose frequencies in every direction
are similar ($\eta_{x}=1.2,\eta_{y}=1.1$) and quite different ($\eta_{x}=5,\eta_{y}=3$)
with each other, respectively.

\begin{figure}[t]
\centering \includegraphics[width=9cm]{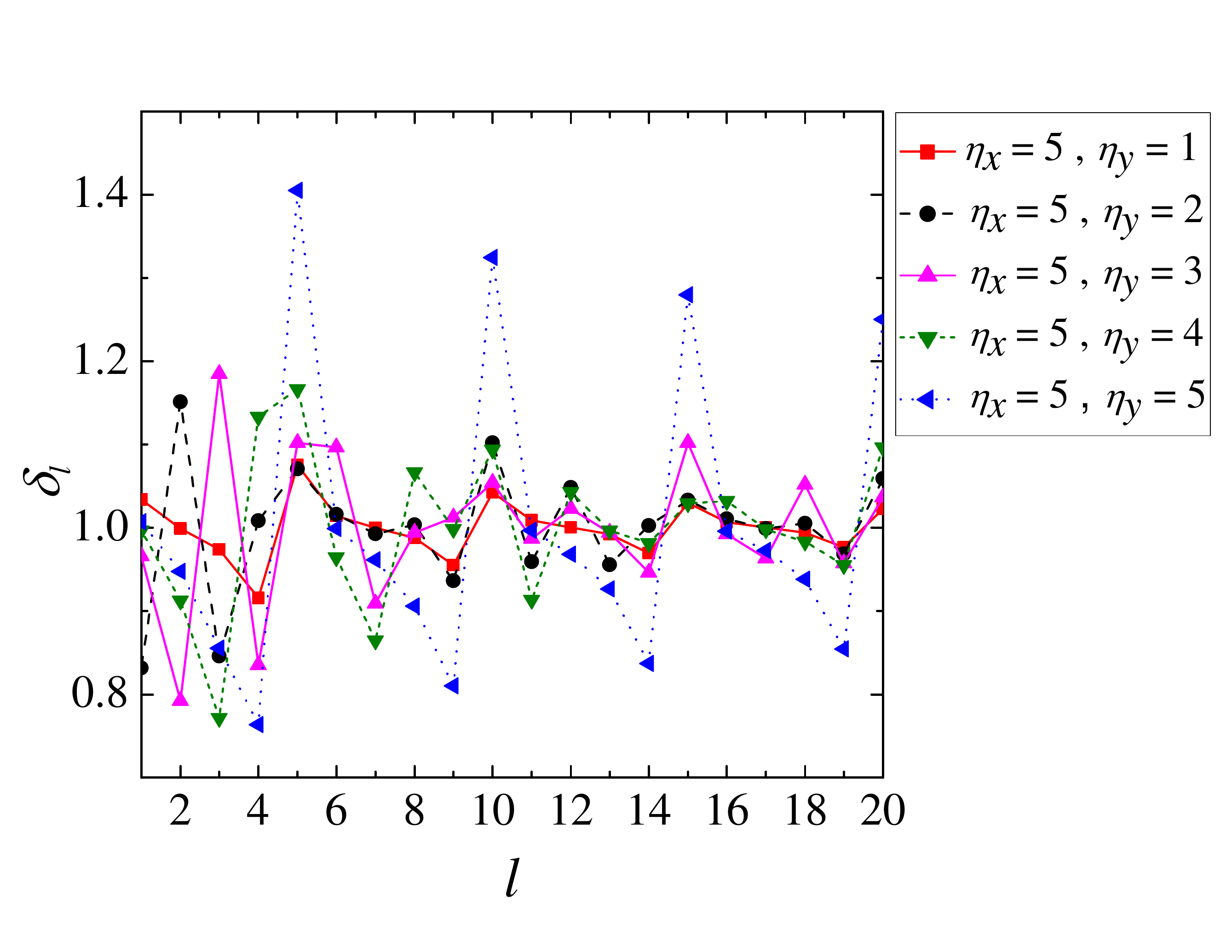}  
\caption{(color online) The normalized energy spacing $\delta_l$ defined in Eq. (\ref{dll})
for the lowest 20 excited states of the systems with $a_{{\rm 3D}}=\infty$, $\eta_x=5$ and $\eta_y$ ($\eta_y=1,2,3,4,5$). Here we also use the natural unit $\hbar=2\mu=\omega_z=1$}
\label{flu} 
\end{figure}


As an example of the application of our method, we make a simple investigation for  the energy-level distribution for the systems with integer aspect ratios $\eta_x$ and $\eta_y$. This property is crucial for many important problems such as thermalization and quantum chaotic behaviors. It is clear that for non-interacting cases (i.e., for $a_{{\rm 3D}}=0$) the energy levels of 
these systems are distributed with equal spacing. However, when $a_{{\rm 3D}}\neq0$
the energy-level distributions become uneven, i.e., the energy spacings become unequal with each other. Nevertheless, for systems with different aspect ratios, the ``degree of unevenness" of the energy-level distributions, or the fluctuation of the energy spacings, are quite different. 
In Ref.~\cite{calarco} it was found that for the  
pancake-shape traps with $\eta_{x}\gg 1$ and $\eta_{y}=1$, the energies are almost distributed
with equal spacing (i.e., the level spacings are almost equal), while for the  cigar-shape traps with $\eta_{x}=\eta_{y}\gg 1$ the eigen-energies are distributed
very even (i.e., the energy spacings fluctuates significantly). Here we study  the intermediate cases between the ``pancake limit" and the ``cigar limit", i.e., the systems with $a_{{\rm 3D}}=\infty$, $\eta_x=5$ and various values of $\eta_y$ ($\eta_y=1,2,3,4,5$).
We define 
$\Delta_{l}$ ($l=1,2,3,...; \Delta_{l}>0$) as the 
spacing between the $l$-th and $(l+1)$-th
excited-state energy of our systems, and
calculate the ``normalized" energy spacing
\begin{eqnarray}
\delta_{l}\equiv\frac{\Delta_{l}}{\frac{1}{20}\sum_{l=1}^{20}\Delta_{l}},\ \ \ (l=1,...,20)\label{dll}
\end{eqnarray}
for the lowest 20 excited states for each system. As shown in Fig.~\ref{flu}, when the system is crossed from the pancake-shape limit
to the cigar-shape, i.e., $\eta_{y}$ is increased from $1$ to $5$,
the fluctuation of $\delta_{l}$ 
monotonically increases, i.e., the energy-level distribution becomes more and more uneven. In future works we will perform more systematic investigations for the energy-level distribution  of two ultracold atoms in a completely anisotropic harmonic trap.

\subsection{Eigen-State of $\hat{H}$}

Furthermore, Eqs. (\ref{eq2}) and (\ref{green0}) yields that for a given eigen-energy $E$ of  the total Hamiltonian $\hat{H}$ for the two-atom relative motion, the corresponding the eigen-state $|\Psi\rangle$ satisfies 
\begin{eqnarray}
\langle{\bf r}|\Psi\rangle&\propto&\langle{\bf r}|\frac{1}{E-\hat{H}_{0}}|{\bf 0}\rangle=\sum_{\bf n}\frac{\langle{\bf r}|{\bf n}\rangle\langle {\bf n}|{\bf 0}\rangle}{E-E_{\bf n}}.
\end{eqnarray}
Thus, considering the normalization condition $\langle \Psi|\Psi\rangle=1$, we obtain the expression for wave function of the eigen-state:
\begin{eqnarray}
\langle{\bf r}|\Psi\rangle  =  \frac{1}{\sqrt{\sum_{{\bf n}}\left|c_{{\bf n}}\right|^{2}}}\sum_{\bf n}c_{\bf n}\phi_{n_x}(\eta_x,x)\phi_{n_y}(\eta_y,y)\phi_{n_z}(\eta_z,z),\nonumber\\
\label{cnn}
\end{eqnarray}
with
\begin{eqnarray}
c_{{\bf n}} =  \frac{1}{(E-E_{{\bf n}})}\phi_{n_x}(\eta_x,0)\phi_{n_y}(\eta_y,0)\phi_{n_z}(\eta_z,0).\label{cnn2}
\end{eqnarray}
Here $(x,y,z)$ are the components of ${\bf r}$, the function $\phi_{n}(\eta,X)$ is defined in Eq. (\ref{phin}). It is just the eigen wave function of a one-dimensional harmonic oscillator and satisfies $\langle {\bf r}|{\bf n}\rangle=\phi_{n_x}(\eta_x,x)\phi_{n_y}(\eta_y,y)\phi_{n_z}(\eta_z,z)$. 
Using Eqs. (\ref{cnn}, \ref{cnn2}) one can easily derive the normalized eigen-state $|\Psi\rangle$ of  $\hat{H}$  from the eigen-energy $E$ \cite{busch,calarco2,guan2b}.


\section{2D Systems}

Now we consider two atoms  confined in a 2D anisotropic harmonic trap in the $x-z$ plane. 
As above, here we assume the  trapping frequencies $\omega_x$ ($\omega_z$)
 in the $x(z)$-direction are same for each atom. Thus, by separating out the CoM degree
of freedom, we can obtain the free Hamiltoian operator for the relative
motion. In our natural unit with $\hbar=2\mu=\omega_z=1$ this free Hamiltonian can be expressed as
\begin{equation}
\hat{H}_{0}^{\rm (2D)}=\hat{{\bf p}}_{\rm 2D}^{2}+\frac{1}{4}\left(\eta_{x}\hat{x}^{2}+\hat{z}^{2}\right),\label{h02d}
\end{equation}
where  $\eta_x=\omega_x/\omega_z$ is the aspect ratio, as in Sec. II. In Eq.~(\ref{h02d}), $\hat{{\bf p}}_{\rm 2D}$ and $\hat{\bm{\rho}}\equiv(\hat{x},\hat{z})$
are the relative momentum and coordinate operators, respectively.
We further assume that the two atoms experience a 2D $s$-wave zero-range
interaction with 2D scattering length $a_{{\rm 2D}}$, and model this potential with the 2D Bethe-Perierls boundary condition
(BPC). Explicitly, in our calculation the eigen-energy $E$ and corresponding
eigen-state $|\Psi\rangle$ for these two interacting atoms should
satisfy the equation 
\begin{equation}
\langle\bm{\rho}|\hat{H}_{0}^{\rm (2D)}|\Psi\rangle=E\langle\bm{\rho}|\Psi\rangle;\ \ {\rm for}\ \rho>0,\label{eigen2}
\end{equation}
and the BPC \cite{2dbpc}
\begin{equation}
\lim_{\rho\rightarrow0}\langle\bm{\rho}|\Psi\rangle\propto\left(\ln\rho-\ln a_{{\rm 2D}}\right),\label{bpc}
\end{equation}
where $|\bm{\rho}\rangle$ is the eigen-state of the relative-coordinate
operator $\hat{\bm{\rho}}$, with the corresponding eigen-energy $\bm{\rho}$,
and $\rho=|\bm{\rho}|$. Namely, the wave function $\langle\bm{\rho}|\Psi\rangle$
satisfies the eigen-equaiton of the free Hamiltonian $\hat{H}_{0}^{\rm (2D)}$
in the region other than the origin (i.e., the region with $\rho>0$),
and has the singular behavior (\ref{bpc}) in the limit $\rho\rightarrow0$,
which describes the interaction effect. As in the 3D cases, this zero-rang interaction only acts on the states in the subspace with even-parity with respect  to the spatial inversion ${\bm \rho}\rightarrow-{\bm \rho}$, and thus in this work we only consider the eigen-energies in this subspace. In addition, in the long-range
limit $\rho\rightarrow\infty$ the wave function $\langle\bm{\rho}|\Psi\rangle$
should satisfy 
\begin{equation}
\lim_{\rho\rightarrow\infty}\langle\bm{\rho}|\Psi\rangle=0.\label{b22}
\end{equation}

\subsection{Expression of $J_{\rm 2D}(E)$}

The solution to Eq. (\ref{eigen2})
and the long-range condition (\ref{b22}) is proportional to the 2D Green's function, i.e.,
\begin{equation}
\langle\bm{\rho}|\Psi\rangle \propto G_0^{\rm (2D)}(E;{\bf \bm{\rho}})\equiv\langle{\bf \bm{\rho}}|\frac{1}{E-\hat{H}_{0}^{\rm (2D)}}|{\bf 0}\rangle.\label{psi2d}
\end{equation}
Therefore, we can derive the algebrac equation for the eigen-energy
$E$ by matching $G_0^{\rm (2D)}(E;{\bf \bm{\rho}})$ with the
BPC (\ref{bpc}). To this end we need to expand $G_0^{\rm (2D)}(E;{\bf \bm{\rho}})$
in the limit $\rho\rightarrow0$. This expansion can be done with
the similar approach as in Sec. II, and we show the detail in Appendix
C. Finally we obtain 
\begin{widetext}
\begin{equation}
\lim_{\rho\rightarrow0}G_0^{\rm (2D)}(E;{\bf \bm{\rho}})=\frac{1}{2\pi}\ln\rho-2\left[W_{\rm 2D}(E)+\int_{0}^{+\infty}I_{\rm 2D}(E,\beta)d\beta\right],\label{g2d}
\end{equation}
where the functions $W_{\rm 2D}(E)$ and $I_{\rm 2D}(E,\beta)$ being defined
as 
\begin{equation}
W_{\rm 2D}(E)=\frac{\sqrt{\eta_{x}}}{2}\sum_{n_{x}\in C_{E}^{\rm (2D)}}\left\{ \frac{\Gamma\left[\frac{1}{4}-\frac{\left(E-\epsilon_{n_{x}}\right)}{2}\right]2^{n_{x}-2}}{\Gamma\left(\frac{1-n_{x}}{2}\right)^{2}\Gamma(n_{x}+1)\Gamma\left[\frac{3}{4}-\frac{\left(E-\epsilon_{n_{x}}\right)}{2}\right]}\right\} -\frac{\gamma}{4\pi}-\frac{1}{8\pi}\ln\left(\frac{\kappa}{4}\right),\label{w2d}
\end{equation}
and 
\begin{eqnarray}
I_{\rm 2D}(E,\beta) & = & \frac{e^{\beta E}}{2}\prod_{\alpha=x,z}\sqrt{\frac{\eta_{\alpha}}{4\pi\sinh\left(\eta_{\alpha}\beta\right)}}-\frac{1}{8\pi\beta}e^{-\kappa\beta}
 -\sqrt{\frac{\eta_{x}}{4\sinh\beta}}\sum_{n_{x}\in C_{E}^{\rm (2D)}}\left\{ \frac{2^{n_{x}-\frac{3}{2}}}{\Gamma\left(\frac{1-n_{x}}{2}\right)^{2}\Gamma(n_{x}+1)}e^{\beta(E-\epsilon_{n_{x}})}\right\} ,\label{iee2d}
\end{eqnarray}
respectively, in our natural unit. Here $\gamma=0.5772...$ is the Euler's constant, $\eta_z=1$, $\epsilon_{n_{x}}=(n_x+1/2)\eta_x$, the parameter $\kappa$ could
be any positive number and the result is independent
of the value of $\kappa$. In Eq. (\ref{iee2d})  $C_{E}^{\rm (2D)}$ is a number set defined
as 
\begin{equation}
C_{E}^{\rm (2D)}:\left\{ \left.n_{x}\right|n_{x}=0,2,4,6,...;\ \epsilon_{n_{x}}+\frac{1}{2}\leq E\right\} .\label{ce2d}
\end{equation}
Clearly, for $E<E_0$ the set $C_{E}^{\rm (2D)}$ is empty and the summation $\sum_{n_{x}\in C_{E}^{\rm (2D)}}$ in the expressions (\ref{w2d}, \ref{iee2d}) of $W_{\rm 2D}$ and $I_{\rm 2D}$ becomes zero.

Combining Eq. (\ref{g2d}), Eq. (\ref{psi2d}) and Eq. (\ref{bpc}),
we obtain the equation for the eigen-energy $E$ of our 2D system,
which has the form of Eq. (\ref{2de}): 
\begin{equation}
J_{{\rm 2D}}(E)=\ln a_{{\rm 2D}}.\label{jj3d-1}
\end{equation}
Here the function $J_{{\rm 2D}}(E)$ is given by
\begin{equation}
J_{{\rm 2D}}(E)=4\pi\left[W_{\rm 2D}(E)+\int_{0}^{+\infty}I_{\rm 2D}(E,\beta)d\beta\right],\label{j2de}
\end{equation}
with $W_{\rm 2D}(E)$ and $I_{\rm 2D}(E,\beta)$ being defined in Eq. (\ref{w2d})
and Eq. (\ref{iee2d}), respectively. One can derive the eigen-energies of the relative motion of these
two atoms by solving Eq. (\ref{jj3d-1}), and further obtain the corresponding
eigen-states.

As in the 3D cases, the summations in Eq. (\ref{w2d}, \ref{iee2d}) are done for finite terms, and the integration in Eq. (\ref{j2de}) converges. Therefore, using Eqs.~(\ref{j2de}, \ref{w2d}, \ref{iee2d}) one can efficiently  calculate $J_{{\rm 2D}}(E)$. 
In addition, for the systems with a 2D isotropic trap ($\omega_x=\omega_z$ or $\eta_x=1$), the equaiton (\ref{jj3d-1}) for the eigen-energy and 
the expression (\ref{j2de}) for the function $J_{\rm 2D}(E)$ are equivalent to the results derived by Ref. \cite{busch}. Explicitly, we have \cite{aa2d}
\begin{eqnarray}
J_{\rm 2D}(E)=\frac{d}{dz}\ln\left[\Gamma(z)\right]\Big \vert_{z=\frac{1-E}2}-\gamma+\frac{1}{2}\ln 2,\ \ ({\rm for}\ \omega_x=\omega_z).
\end{eqnarray}

\subsection{Techniques for Fast Calculation of $J_{{\rm 2D}}(E)$}

The techniques shown in Sec. II.B for the fast calculation of  $J_{{\rm 3D}}(E)$ can also be directly generalized to the 2D case (Appendix D). In particular, as proved in Appendix D,  $J_{\rm 2D}(E)$ given by Eq. (\ref{j2de}) can be re-expressed as
\begin{eqnarray}
J_{{\rm 2D}}(E) & = & 4\pi\left[\int_{0}^{\Lambda}A_{{\rm 2D}}(E,\beta)d\beta+B^{(1)}_{\rm 2D}(E, \Lambda)
 +B^{(2)}_{\rm 2D}(E,\Lambda)-\frac{\Gamma(0,\kappa\Lambda)}{8\pi}\right],
\label{dg2d22}
\end{eqnarray}
where $\Lambda$ and $\kappa$ are arbitrary finite positive numbers, as in Eqs. (\ref{w2d}, \ref{iee2d}, \ref{dg22}), $\Gamma[a,z]$ is the incomplete Gamma function and the functions $A_{{\rm 2D}}(E,\beta)$ and $B^{(1,2)}_{{\rm 2D}}(E,\Lambda)$ are defined as
\begin{eqnarray}
A_{{\rm 2D}}(E,\beta)&=& \frac{e^{\beta E}}{2}\prod_{\alpha=x,z}\sqrt{\frac{\eta_{\alpha}}{4\pi\sinh\left(\eta_{\alpha}\beta\right)}}-\frac{1}{8\pi\beta}e^{-\kappa\beta};
\label{a2d}\\
\nonumber\\
B^{(1)}_{{\rm 2D}}(E,\Lambda)&=&-\frac{\gamma}{4\pi}-\frac{1}{8\pi}\ln\left(\frac{\kappa}{4}\right)
+\sum_{n_{x}\in C_{E}^{{\rm (2D)}}}\left\{
\frac{\sqrt{\eta_{x}} 2^{n_{x}-3}e^{(E-\epsilon_{n_x}-{3\over2})\Lambda}\sqrt{e^{2\Lambda}-1}\cdot\Gamma\left({1\over4}-{{E-\epsilon_{n_{x}}}\over2}\right)}
{\Gamma\left(\frac{1-n_{x}}{2}\right)^{2}\Gamma(n_{x}+1)\Gamma\left({5\over4}-{{E-\epsilon_{n_{x}}}\over2}\right)}\right.  \nonumber\\
&&\ \ \ \ \ \ \ \ \ \ \ \ \ \ \ \ \ \ \ \ \ \ \ \ \ \ \ \ \ \ \ \ \ \ \ \ \ \ \ \ \ \
\left.
\times _{2}F^{1}\qty[1,\frac{3}{4}-\frac{E-\epsilon_{n_x}}{2},\frac{5}{4}-\frac{E-\epsilon_{n_x}}{2},e^{-2\Lambda}]
\begin{array}{c}
\vspace{10pt}
\end{array}
\right\};
\label{b2d1}
\end{eqnarray}
and
\begin{eqnarray}
B^{(2)}_{{\rm 2D}}(E,\Lambda)&=&-\sum_{n_x\notin
C_{E}^{{\rm (2D)}}
}\frac{2^{n_{x}-\frac{3}{2}}\cdot\sqrt{\eta_{x}\csch{\Lambda}}}{\Gamma\left(\frac{1-n_{x}}{2}\right)^{2}\Gamma(n_{x}+1)} \cdot
\frac{e^{(E-\epsilon_{n_x}-2)\Lambda}(e^{2\Lambda}-1)\times _{2}\! F^{1}\qty[1,\frac{3}{4}-\frac{E-\epsilon_{n_x}}{2},\frac{5}{4}-\frac{E-\epsilon_{n_x}}{2},
e^{-2\Lambda}]}
{2\qty(E-\epsilon_{n_x})-1},
 \nonumber\\
 \label{b2d2}
\end{eqnarray}
respectively, with $ _{2}F^{1}$ being the Hypergeomtric function. As in the 3D cases, the
expression (\ref{dg2d22})  of $J_{\rm 2D}(E)$ has  the advantages (A) and (B)
shown in Sec. II. B. Therefore, the numerical calculations based on Eq.  (\ref{dg2d22}) is quite possibly to be faster then the one based on Eq.  (\ref{j2de}), especially for the high-energy cases with large $E$.

\end{widetext}

\begin{figure*}[t]
\centering \includegraphics[width=6cm]{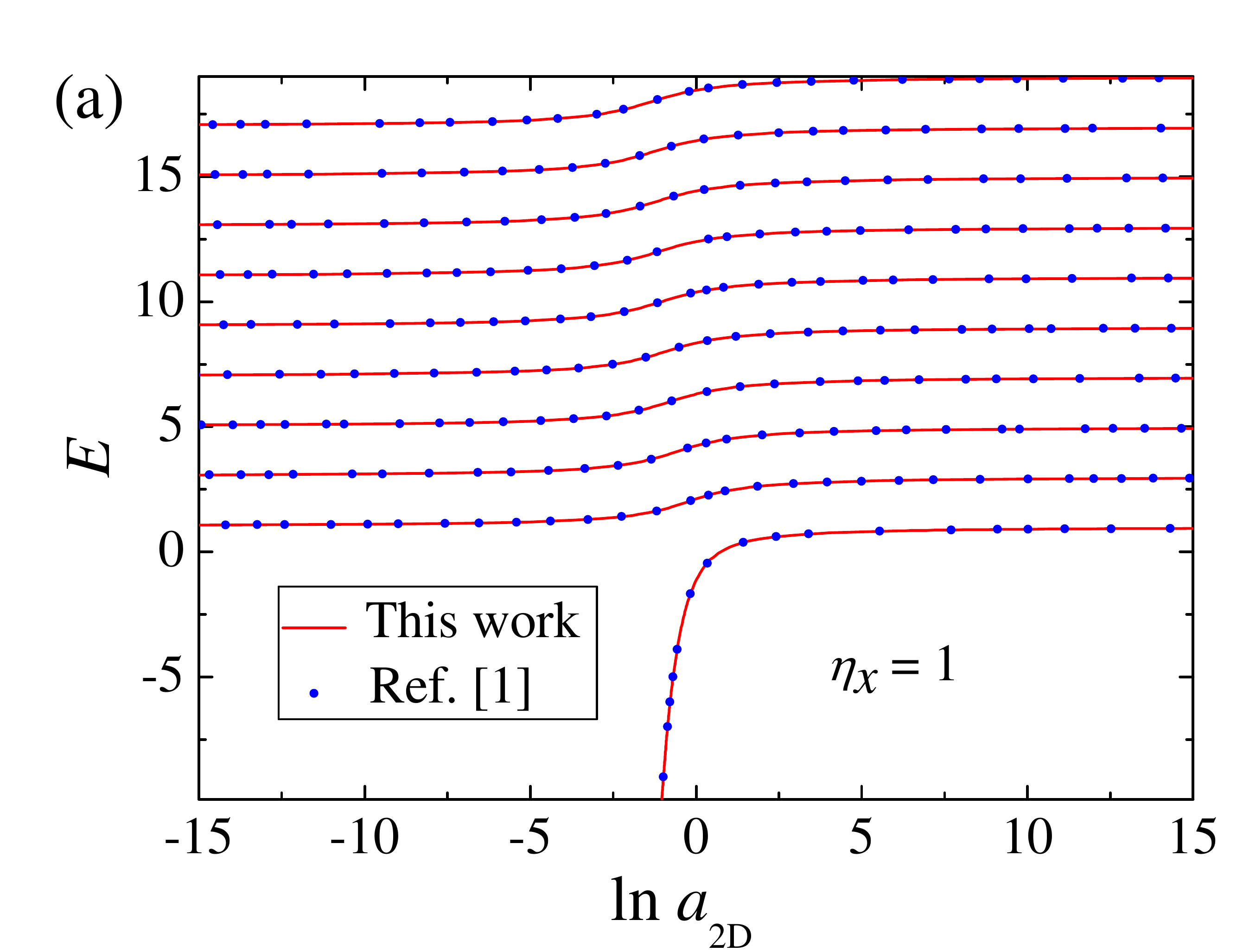}  \includegraphics[width=6cm]{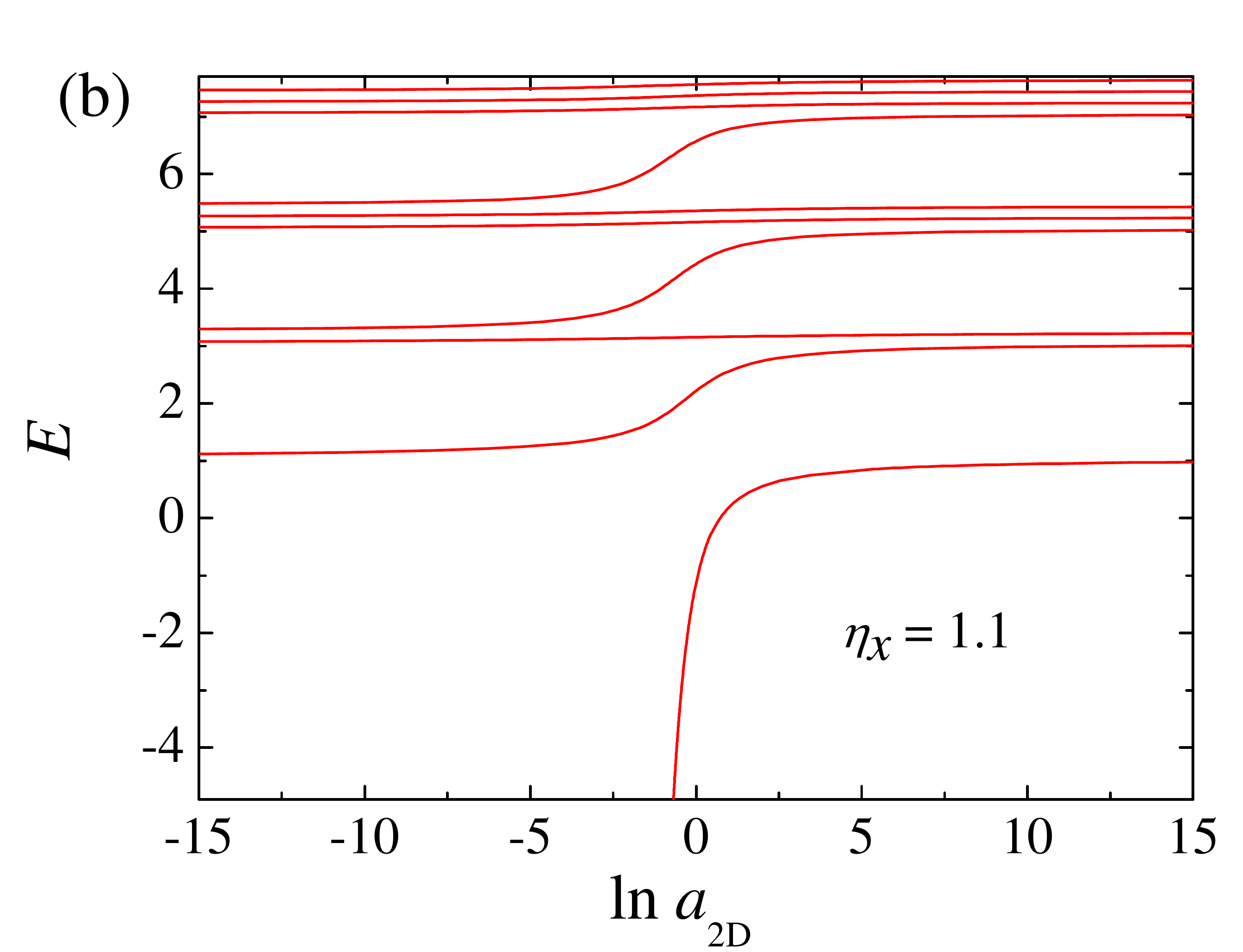} 
\includegraphics[width=6cm]{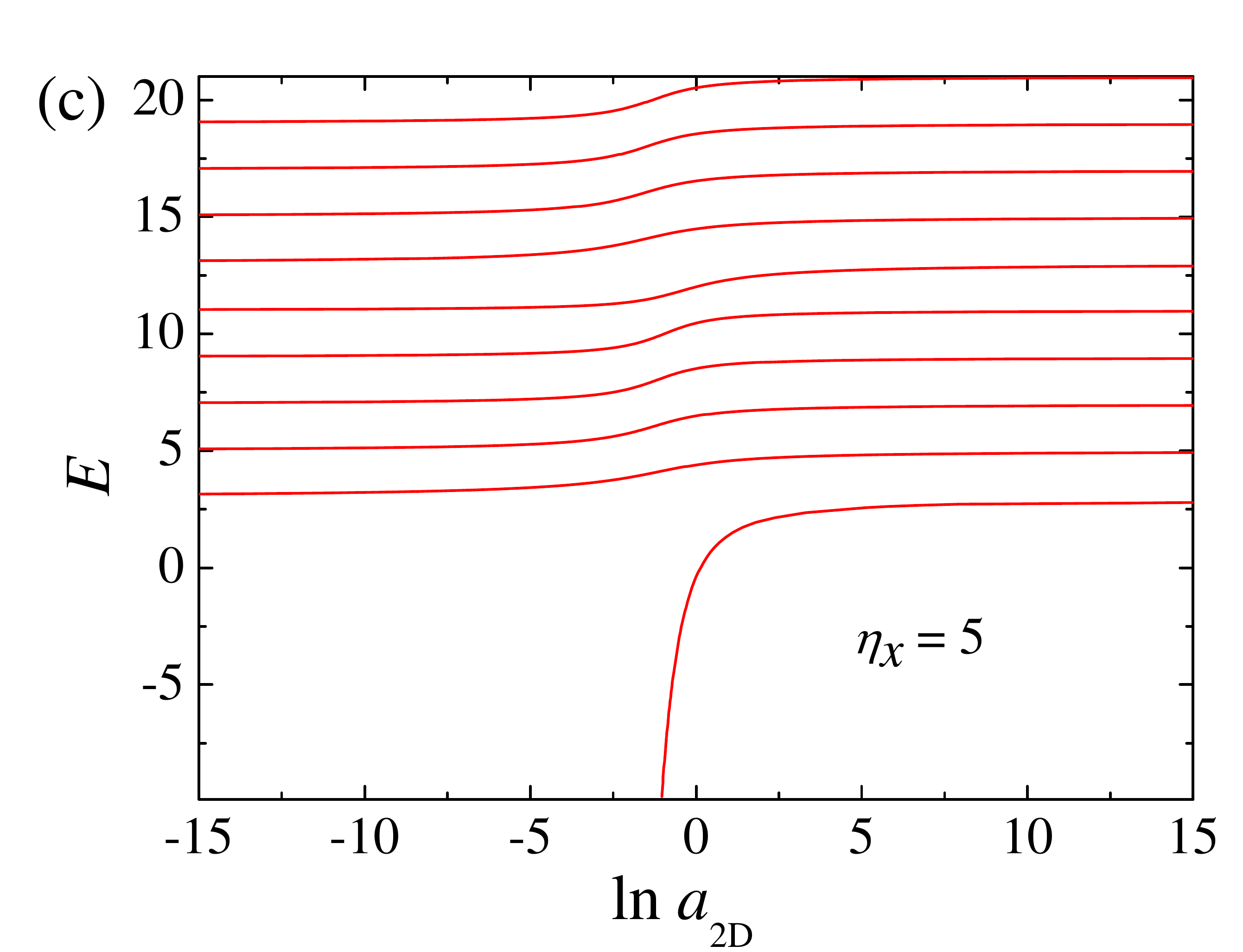} 
 \caption{(color online)  The energy spectrum of the relative motion of two atoms in a 2D harmonic trap with $\eta_x=1$ (a),
$\eta_x=1.1$ (b) and $\eta_x=5$ (c). We show the results given by Eq. (\ref{2de}) and the expression (\ref{j2de}) of the function $J_{\rm 2D}(E)$ with solid lines. 
In (a) we also show the results given by Ref. \cite{busch}  with blue dots. Here we use the natural unit $\hbar=2\mu=\omega_z=1$.}
\label{2d} 
\end{figure*}

\subsection{Energy Sepectrum and Eigen-States}

In Fig.~\ref{2d} (a) we illustrate the energy
spectrum for the cases with an isotropic 2D trap ($\eta_{x}=1$),
and show that our results are same as the ones from Ref. \cite{busch}.
In Fig.~\ref{2d} (b) and (c) we show the results for anisotropic
2D traps which have similar ($\eta_{x}=1.1$) and quite different
($\eta_{x}=5$) frequencies in the $x$- and $z$-directions.

In addition, similar to Sec. II.D, one can also derive the  corresponding eigen-state $|\Psi\rangle$ with eigen-energy $E$, which satisfies Eq. (\ref{eigen2}) and the boundary conditions (\ref{bpc}, \ref{b22}), as well as the normalization condition $\langle \Psi|\Psi\rangle=1$. With Eq. (\ref{psi2d}) and the similar approach as in Sec. II. D we obtain
\begin{eqnarray}
\langle{\bm \rho}|\Psi\rangle  =  \frac{1}{\sqrt{\sum_{\bf m}\left|d_{{\bf m}}\right|^{2}}}\sum_{\bf m}d_{\bf m}\phi_{m_x}(\eta_x,x)\phi_{m_z}(1,z),\nonumber\\
\label{cnn}
\end{eqnarray}
with
\begin{eqnarray}
d_{{\bf m}} =  \frac{1}{\left(E-E^{\rm (2D)}_{{\bf m}}\right)}\phi_{m_x}(\eta_x,0)\phi_{m_z}(1,0).\label{cnn2}
\end{eqnarray}
Here $(x,z)$ are the components of ${\bm \rho}$, ${\bf m}=(m_x,m_y)$ with $m_{x,y}=0,1,2,...$, $E^{\rm (2D)}_{{\bf m}}=(m_x+1/2)\eta_x+(m_z+1/2)$, and the function $\phi_{n}(\eta,X)$ is defined in Eq. (\ref{phin}).

\section{generalization of our approach to other problems}

At the end of this work, we briefly summarize the
main ideas of our  approach used in the above calculations, and discuss how to generalize these ideas. Here we take the 3D systems as an example. 

The key step for solving the two-body problem with zero-rang interaction is the calculation of the free Green's operator ${\hat G}_0(E)\equiv1/[E-\hat{H}_{0}]$. When $E$ is less than the ground-state energy $E_0$ of $\hat{H}_{0}$, ${\hat G}_0(E)$ can be directly expressed as the Laplace transform of of the imaginary-time
evolution operator $e^{-\beta\hat{H}_{0}}$, i.e., we have ${\hat G}_0(E)=-\int_{0}^{+\infty}e^{\beta(E-\hat{H}_{0})}d\beta$. However, when $E>E_0$ this integration diverges. 
To solve this problem, we separate
${\hat G}_0(E)$ into two parts, i.e., the contributions
from the high-energy states with ${\bf n}\in U_{E}$ and the ones from the low-energy states with ${\bf n}\in L_{E}$, as shown in Eq. (\ref{sep}).
Furthermore, as shown in Eqs. (\ref{g2}, \ref{ss2}), the first part
can still be expressed as the Laplace transform
of the operator $e^{-\beta\hat{H}_{0}^{{\rm (3D)}}}-\sum_{{\bf n}\in L_{E}}|{\bf n}\rangle\langle{\bf n}|e^{-\beta\left(E_{{\bf n}}\right)}$,
which converges for for any non-zero $z$, and the second part only includes finite terms. 

It is pointed out that, the
definitions of the sets $L_{E}$ and $U_{E}$ are not unique. The
only requirements are:
\begin{itemize}

\item[(i)]: $L_{E}$ is the complement of $U_{E}$. 

\item[(ii)]: $E_{{\bf n}}>E$ for $\forall {\bf n}\in U_{E}$. Namely, the set $L_{E}$ includes (but is not limited to) all ${\bf n}$ which satisfies $E_{{\bf n}}<E$.
\end{itemize}
For instance,
an alternative definition of these two sets is: $U_{E}:\{{\bf n}|E_{{\bf n}}>E\}$
and $L_{E}:\{{\bf n}|E_{{\bf n}}\leq E\}$. 

Using this method we can derive the helpful expression (\ref{g3}) of  Green's function $G^{(0)}(E,z{\bf e}_{z})$, which is just the matrix element $\langle z{\bf e}_{z}|{\hat G}_0(E)|{\bf 0}\rangle$ of the free Green's operator ${\hat G}_0(E)$. In this expression there is just a summation for finite terms and a one-dimensional  integration $\int_{0}^{+\infty}d\beta\left[K(z;E,\beta)-F(z;E,\beta)\right]$ which converges for any finite $z$.

To complete the calculation we still require to remove the divergent term of the above integration in the limit $z\rightarrow 0$. This divergent term is contributed by the leading term $e^{-\frac{z^{2}}{4\beta}}/(4\pi\beta)^{\frac{3}{2}}$ of the integrand $K(z;E,\beta)-F(z;E,\beta)$ in the limit $\beta\rightarrow 0^+$. Therefore, we can remove it 
via the technique used
in Eqs. (\ref{int3a}, \ref{kp}, \ref{gexpa}).

Our above approach for the calculation of the two-body free Green's function can be directly generalized to other few-body  problems of ultracold
atoms, especially the ones where  the analytical expressions of the eigen-states and 
imaginary-time propagator of the free Hamiltonian are known, {\it e.g.}, the few-body problems in mixed-dimensional systems \cite{shina}. Here we emphasize    that, with the help of the Laplace transformation for imaginary-time propagator, the free Green's operator
can always be expressed as a one-dimensional integration, no matter how
many degrees of freedoms are involved in the system. Thus, the free Green's function given by our method always includes a converged one-dimensional integration and summations for finite terms.

\section{Summary}

In this work we derive the algebraic equations for the eigen-energies
of two atoms in a 2D or 3D harmonic trap. Our results is applicable for general cases,
no matter if the trap is completely anisotropic or has spherical or
axial symmetry. Using our results one can easily derive the complete
energy spectrum, which can be used for the further theoretical or
experimental studies of dynamical or thermodynamical problems. Our approach can be used in other few-body problems of confined ultracold
atoms.

{{\it Note added:} When we finished this work, we realized that recently there is a related work 
\cite{bougas2020stationary}. The authors derived the expression of  $J_{\rm 2D}(E)$ for $E<E_0$, and a recurrence relation of $J_{\rm 2D}(E)$ for arbitrary $E$. With this recurrence relation they also obtained the complete energy spectrum of two atoms in a 2D anisotropic confinement, as well as the eigen-states.}

\begin{acknowledgments}
This work is supported by the National Key R$\&$D Program of China
(Grant No. 2018YFA0306502 (PZ), 2018YFA0307601 (RZ)), NSFC Grant No.
11804268 (RZ), 11434011(PZ), 11674393(PZ), as well as the Research
Funds of Renmin University of China under Grant No. 16XNLQ03(PZ). 
\end{acknowledgments}

\appendix

\begin{widetext}
\section{Proof of Eqs. (\ref{wee}) and (\ref{iee})}

In this appendix we prove Eqs. (\ref{wee}) and (\ref{iee}) in Sec. II.A. To this end, we first show some results on 1D harmonic oscillator, which will be used in our calculation.

\subsection{Some properties of 1D harmonic oscillator} 

Let us consider a 1D harmonic oscillator with frequency $\eta$ and mass $\mu$. The Hamiltonian of this oscillator is ($\hbar=2\mu=1$)
\begin{eqnarray}
{\hat H}_{\rm ho}={\hat P}^2+\frac{\eta^2{\hat X}^2}{4},
\end{eqnarray}
with ${\hat X}$ and  ${\hat P}$ being the coordinate and momentum operator, respectively. The eigen-energy of  ${\hat H}_{\rm ho}$ is 
\begin{eqnarray}
E_n=\left(n+\frac{1}{2}\right)\eta;\ \ \ (n=0,1,2,...), 
\end{eqnarray}
and the wave function of the eigen-state $|n\rangle$ corresponding to $E_n$ can be expressed as
\begin{eqnarray}
\langle X|n\rangle\equiv\phi_n(\eta,X)=\left(\frac{\eta}{2\pi}\right)^{\frac 14}\frac{e^{-\frac{\eta X^2}4}}{\sqrt{2^{n}\Gamma(n+1)}}H_{n}\left(\sqrt{\frac{\eta}{2}}X\right),\label{phin}
\end{eqnarray}
with $|X\rangle$ being the eigen-state of the position operator ${\hat X}$ with eigen-value $X$, $H_n(X)$ and $\Gamma(\alpha)$ being the Hermitian polynomial and the Gamma function, respectively. The wave function $\phi_n(\eta,X)$ also satisfies
\begin{eqnarray}
\langle X|e^{-\beta {\hat H}_{\rm ho}}|0\rangle=\sum_{n=0}^{+\infty}\phi_{n}(\eta,X)\phi_{n}^\ast(\eta,0)
e^{-\beta (n+1/2)\eta}  =  \sqrt{\frac{\eta}{4\pi\sinh(\eta\beta)}}\exp\left[-\frac{\eta X^{2}\coth{(\eta\beta)}}{4}\right]\label{p1dd}
\end{eqnarray}
for $\beta>0$.

Now we consider the Green's function $g(\xi;\eta;X)$ of the 1D harmonic oscillator, which is defined as
\begin{eqnarray}
g(\xi;\eta;X)\equiv\langle X|\frac{1}{\xi-{\hat H}_{\rm ho}}|0\rangle=\sum_{n=0}^{+\infty}\frac{\phi_{n}(\eta,X)\phi_{n}^\ast(\eta,0)}{\xi-E_n}.
\end{eqnarray}
This function satisfies the differential equation 
\begin{eqnarray}
\xi \cdot g(\xi;\eta;X) +\frac{d^2}{dX^2}g(\xi;\eta;X)-\frac{\eta^2X^2}{4}g(\xi;\eta;X)=\delta(X)\label{deg}
\end{eqnarray}
and the boundary condition
\begin{eqnarray}
\lim_{|X|\rightarrow\infty}g(\xi;\eta;X)=0.\label{bbd}
\end{eqnarray}
To derive $g(\xi;\eta;X)$,
we can first solve the equation (\ref{deg}) in the regions $X>0$ and $X<0$ with the boundary condition (\ref{bbd}), and then match the solution with the connection condition at $X=0$, which is given by the term $\delta(X)$ in Eq. (\ref{deg}). With this approach we obtain
\begin{eqnarray}
g(\xi;\eta;X)=\sum_{n=0}^{+\infty}\frac{\phi_{n}(\eta,X)\phi_{n}^\ast(\eta,0)}{\xi-E_n}= -\frac{\Gamma(\frac{1}{4}-\frac{\xi}{2\eta})}{2^{\frac{5}{4}+\frac{\xi}{2\eta}}\sqrt{\pi\eta}}D_{\frac{\xi-\eta/2}{\eta}}\left(\sqrt{\eta}X\right),\label{g1d}
\end{eqnarray}
where $D_\lambda(\alpha)$ is the parabolic cylinder function. Eq. (\ref{g1d}) and the property of the parabolic cylinder function further yields
\begin{eqnarray}
g(\xi;\eta;0)=\sum_{n=0}^{+\infty}\frac{|\phi_{n}(\eta,0)|^2}{\xi-(n+1/2)\eta}{= -\frac{\Gamma(\frac{1}{4}-\frac{\xi}{2\eta})}
{2\sqrt{2\eta}\Gamma(\frac{3}{4}-\frac{\xi}{2\eta})}.}
\label{g1d2}
\end{eqnarray}

\subsection{Proof of the two equations}

Now we prove Eq.~(\ref{wee}) and Eq.~(\ref{iee}) in our main text, which are about the expressions of the functions $W_{{\rm 3D}}(E)$ and $I_{{\rm 3D}}(E,\beta)$, respectively. As shown in Sec. II.A, these two functions are defined as
\begin{eqnarray}
W_{{\rm 3D}}(E)\equiv Q(0,E,\beta),\label{www}
\end{eqnarray}
and
\begin{eqnarray}
I_{{\rm 3D}}(E,\beta)\equiv-\tilde{K}(0;E,\beta)+F(0;E,\beta),\label{iii}
\end{eqnarray}
with the functions $Q(z,E,\beta)$, $\tilde{K}(z;E,\beta)$ defined in  Eq. (\ref{fr}) and Eq. (\ref{kp}), respectively.  Thus, we have
\begin{eqnarray}
Q(0;E,\beta)  &=&  \sum_{{\bf n}\in L_{E}}\frac{\langle{\bf 0}|{\bf n}\rangle\langle{\bf n}|{\bf 0}\rangle}{E-E_{{\bf n}}};\label{qq0}\\
F(0;E,\beta) & = & \sum_{{\bf n}\in L_{E}}\langle{\bf 0}|{\bf n}\rangle\langle{\bf n}|{\bf 0}\rangle e^{-\beta\left(E_{{\bf n}}-E\right)}.\label{ff0}
\end{eqnarray}
It is clear that the eigen-state $|{\bf n}\rangle$ of the 3D free Hamiltonian ${\hat H}_0$, which is defined in Sec. II.A, satisfies 
$
\langle{\bf 0}|{\bf n}\rangle=0$ 
when $n_x$ or $n_y$ is odd.
Using this fact and the definitions of the sets $L_E$ and $C_E^{\rm (3D)}$, which are given in Eqs. (\ref{ne}, \ref{lep}) of our main text, we obtain
\begin{eqnarray}
Q(0;E,\beta)  &=&  \sum_{(n_{x},n_{y})\in C_{E}^{\rm (3D)}}\sum_{n_{z}=0}^{+\infty}\frac{\langle{\bf 0}|{\bf n}\rangle\langle{\bf n}|{\bf 0}\rangle}{E-E_{{\bf n}}}\nonumber\\
&&=  \sum_{(n_{x},n_{y})\in C_{E}^{\rm (3D)}}
\left\{
|\phi_{n_x}(\eta_x,0)|^2|\phi_{n_y}(\eta_y,0)|^2
\left[\sum_{n_{z}=0}^{+\infty}
\frac{
|\phi_{n_z}(\eta_z,0)|^2
}{\left(E-\epsilon_{n_{x}}-\epsilon_{n_{y}}\right)-\epsilon_{n_{z}}}\right]
\right\},\label{wwp}
\end{eqnarray}
and
\begin{eqnarray}
F(0;E,\beta)  &=&  \sum_{(n_{x},n_{y})\in C_{E}^{\rm (3D)}}\sum_{n_{z}=0}^{+\infty}\langle{\bf 0}|{\bf n}\rangle\langle{\bf n}|{\bf 0}\rangle e^{-\beta\left(E_{{\bf n}}-E\right)}
\nonumber\\
&&=  \sum_{(n_{x},n_{y})\in C_{E}^{\rm (3D)}}
\left\{
|\phi_{n_x}(\eta_x,0)|^2|\phi_{n_y}(\eta_y,0)|^2
e^{\beta(E-\epsilon_{n_{x}}-\epsilon_{n_{y}})}
\left[\sum_{n_{z}=0}^{+\infty}
|\phi_{n_z}(\eta_z,0)|^2e^{-\beta\epsilon_{n_{z}}}\right]
\right\},\label{iip}
\end{eqnarray}
where $\eta_{x,y,z}$ and $\epsilon_{n_{x,y,z}}$ are defined in Sec. II, and the function $\phi_n(\eta,X)$ is defined in Eq. (\ref{phin}). 

Substituting Eq. (\ref{g1d2}) into Eq. (\ref{wwp}) and then into Eq. (\ref{www}), and using the property
\begin{eqnarray}
H_{n}(0)=\frac{2^{n}\sqrt{\pi}}{\Gamma(\frac{1-n}{2})}\label{hhh0}
\end{eqnarray}
of the Hermitian polynomial, we can derive Eq. (\ref{wee}). Moreover, Substituting Eq. (\ref{p1dd}) into Eq. (\ref{iip}) and then into Eq. (\ref{iii}), and using Eqs. (\ref{hhh0}), (\ref{kp}) and (\ref{kr}), we can derive Eq. (\ref{iee}).

\section{Proof of Eq. (\ref{dg22}) }
In this appendix we prove Eq. (\ref{dg22})  in Sec. II.B. To this end, we separate the integration in Eq. (\ref{dg}) into two parts, i.e.,
\begin{eqnarray}
\int_{0}^{\infty} I_{{\rm 3D}}(E,\beta)  d\beta=\int_{0}^{\Lambda} I_{{\rm 3D}}(E,\beta)  d\beta+\int_{\Lambda}^{\infty} I_{{\rm 3D}}(E,\beta)  d\beta, \label{bb1}
\end{eqnarray}
with $\Lambda$ being an arbitrary finite positive number.
In addition, using the definition (\ref{iee}) of $I_{{\rm 3D}}(E,\beta)$, we immediately obtain
\begin{align}
&\int_{0}^{\Lambda} I_{{\rm 3D}}(E,\beta)  d\beta\nonumber\\
=&\int_{0}^{\Lambda}A_{{\rm 3D}}(E,\beta)d\beta
+\sum_{(n_{x},n_{y})\in C_{E}^{{\rm (3D)}}}
\int_{0}^{\Lambda}
 \frac{2^{n_{x}+n_{y}-2}\sqrt{\pi\eta_{x}\eta_{y}}\ e^{\beta(E-\epsilon_{n_{x}}-\epsilon_{n_{y}})}}
 {\Gamma\left(\frac{1-n_{x}}{2}\right)^{2}\Gamma\left(\frac{1-n_{y}}{2}\right)^{2}\Gamma(n_{x}+1)\Gamma(n_{y}+1)\sqrt{\sinh\beta}}
d\beta\nonumber\\
=& \int_{0}^{\Lambda}A_{{\rm 3D}}(E,\beta)d\beta\nonumber\\
&+\sum_{(n_{x},n_{y})\in C_{E}^{{\rm (3D)}}}
 \frac{2^{n_{x}+n_{y}-{5\over2}}\sqrt{\pi\eta_{x}\eta_{y}}}
 {\Gamma\left(\frac{1-n_{x}}{2}\right)^{2}\Gamma\left(\frac{1-n_{y}}{2}\right)^{2}\Gamma(n_{x}+1)\Gamma(n_{y}+1)}
 \Bigg\{ {\sqrt{\pi} \Gamma\left({1\over4}-{{E-\epsilon_{n_{x}}-\epsilon_{n_{y}}}\over2}\right)\over  \Gamma\left({3\over4}-{{E-\epsilon_{n_{x}}-\epsilon_{n_{y}}}\over2}\right)}-  
 \nonumber \\
  \nonumber \\
& \ \ \ \ \ \  \ \ \ \ \ \ \ \ \ \ \ \ \ \frac{\Gamma\left({1\over4}-{{E-\epsilon_{n_{x}}-\epsilon_{n_{y}}}\over2}\right)}{\Gamma\left({5\over4}-{{E-\epsilon_{n_{x}}-\epsilon_{n_{y}}}\over2}\right)}e^{(E-\epsilon_{n_x}-\epsilon_{n_y}-{3\over2})\Lambda}\sqrt{e^{2\Lambda}-1}\times _{2}\!F^{1}\qty[1,\frac{3}{4}-\frac{E-\epsilon_{n_x}-\epsilon_{n_y}}{2},\frac{5}{4}-\frac{E-\epsilon_{n_x}-\epsilon_{n_y}}{2},
e^{-2\Lambda}]
 \Bigg\}
 \label{bb2}
\end{align}
with $A_{{\rm 3D}}(E,\beta)$ being defined in Eq. (\ref{a3d}).

Now we calculate the term $\int_{\Lambda}^{\infty} I_{{\rm 3D}}(E,\beta)  d\beta$ in Eq. (\ref{bb1}).
We first notice that, according to Eqs.~(\ref{iii},\ref{iip},\ref{kp},\ref{kr}), $I_{{\rm 3D}}(E,\beta)$ can be re-expressed as
 \begin{eqnarray}
I_{{\rm 3D}}(E,\beta)&=&\left(\frac{1}{4\pi\beta}\right)^{\frac{3}{2}}
- \sum_{n_x,n_y,n_z}\langle{\bf 0}|{\bf n}\rangle\langle{\bf n}|{\bf 0}\rangle e^{-\beta\left(E_{{\bf n}}-E\right)}
+\sum_{(n_{x},n_{y})\in C_{E}^{\rm (3D)}}\sum_{n_{z}=0}^{+\infty}\langle{\bf 0}|{\bf n}\rangle\langle{\bf n}|{\bf 0}\rangle e^{-\beta\left(E_{{\bf n}}-E\right)}\nonumber\\
&=&\left(\frac{1}{4\pi\beta}\right)^{\frac{3}{2}}-
\sum_{(n_{x},n_{y})\notin C_{E}^{\rm (3D)}}\sum_{n_{z}=0}^{+\infty}\langle{\bf 0}|{\bf n}\rangle\langle{\bf n}|{\bf 0}\rangle e^{-\beta\left(E_{{\bf n}}-E\right)}\nonumber\\
&=&  \left(\frac{1}{4\pi\beta}\right)^{\frac{3}{2}}-\sum_{(n_{x},n_{y})\notin C_{E}^{\rm (3D)}}
\left\{
|\phi_{n_x}(\eta_x,0)|^2|\phi_{n_y}(\eta_y,0)|^2
e^{\beta(E-\epsilon_{n_{x}}-\epsilon_{n_{y}})}
\left[\sum_{n_{z}=0}^{+\infty}
|\phi_{n_z}(\eta_z,0)|^2e^{-\beta\epsilon_{n_{z}}}\right]
\right\},\label{b2}
\end{eqnarray}
where $\epsilon_{n_{x,y,z}}$ is defined in Sec. II, and the function $\phi_n(\eta,X)$ is defined in Eq. (\ref{phin}). Moreover, Substituting Eq. (\ref{p1dd}) and Eq.  (\ref{hhh0}) into Eq. (\ref{b2}), we further obtain
\begin{eqnarray}
I_{{\rm 3D}}(E,\beta)
 & = & \left(\frac{1}{4\pi\beta}\right)^{\frac{3}{2}}-\sqrt{\frac{\pi\eta_{x}\eta_{y}}{8\sinh\beta}}\sum_{(n_{x},n_{y})\notin C_{E}^{{\rm (3D)}}}\left\{ \frac{2^{n_{x}+n_{y}-\frac{1}{2}}}{\Gamma\left(\frac{1-n_{x}}{2}\right)^{2}\Gamma\left(\frac{1-n_{y}}{2}\right)^{2}\Gamma(n_{x}+1)\Gamma(n_{y}+1)}e^{\beta(E-\epsilon_{n_{x}}-\epsilon_{n_{y}})}\right\},
 \label{itd}
\end{eqnarray}
where $\eta_{x,y,z}$ is defined in Sec. II.
Doing the integration $\int_{\Lambda}^{\infty} I_{{\rm 3D}}(E,\beta)  d\beta$ in both sides of Eq. (\ref{itd}), we further obtain
\begin{eqnarray}
\int_{\Lambda}^{\infty} I_{{\rm 3D}}(E,\beta)  d\beta
&=&B^{(2)}_{\rm 3D}(E,\Lambda)+\qty(\frac{1}{2\pi})^{3/2}\frac{1}{\sqrt{2\Lambda}},
 \label{itdone}
\end{eqnarray}
where $B^{(2)}_{\rm 3D}(E,\Lambda)$ is defined in Eq. (\ref{b23d}).

Substituting Eqs. (\ref{bb2}, \ref{itdone}) into Eq. (\ref{bb1}) and then into Eq. (\ref{dg}), and further using Eqs. (\ref{wee}), we can derive Eq. (\ref{dg22}).


\section{Proof of Eqs. (\ref{w2d}) and (\ref{iee2d}) }

In this appendix we prove Eqs. (\ref{w2d}) and (\ref{iee2d}) in Sec. III, which is related to the behavior of the 2D Green's function $G_{0}^{\rm (2D)}(E ; \bm{\rho})$ in the limit $|{\bm \rho}|\rightarrow 0$. Our approach is similar to the method used in Sec. II.A. 

As in Sec. II and Appendix A, by choosing ${\bm \rho}=\rho{\bf e}_z$ ($\rho>0$) and making direct calculations we can find that 
\begin{eqnarray} 
G_{0}^{\rm (2D)}(E ; \rho{\bf e}_z)=-\int_{0}^{\infty}\left[K_{\rm 2D}(\rho;E,\beta) -Y(\rho;E,\beta)\right] d\beta
+Z(\rho;E), \label{blim_g2d}
\end{eqnarray}
which is similar to the expression (\ref{g3}) of the 3D free Green's function.
Here the functions $K_{\rm 2D}(\rho;E,\beta)$,  $Y(\rho;E,\beta)$ and $Z(\rho;E)$ are defined as
\begin{align}
&K_{\rm 2D}(\rho;E,\beta)=e^{\beta E}\langle \rho{\bf e}_z|e^{-\beta\hat{H}_{0}^{(2 D)}}| \bm{0}\rangle
=\left(\prod_{\alpha=x, z} \sqrt{\frac{ \eta_{\alpha}}{4 \pi  \sinh \left(\eta_{\alpha} \beta\right)}}\right) \cdot
\exp\left[\beta E-\frac{ \operatorname{coth}\left( \beta\right)}{4}\rho^{2}\right];\label{bx}\\
&Y(\rho;E,\beta)=\sum_{n_{x} \in C_{E}^{(2D)}}\sum_{n_{z}=0}^{+\infty} e^{\beta\left(E-E_{\bf n}\right)}\langle \rho{\bf e}_z | {\bf n}\rangle\langle{\bf n} | {\bf 0}\rangle; \label{by}\\
&Z(\rho;E)=\sum_{n_{x} \in C_{E}^{(2D)}} \sum_{n_{z}=0}^{+\infty}\frac{\langle \rho{\bf e}_z | {\bf n}\rangle\langle{\bf n} | {\bf 0}\rangle}{E- E_{\bf n}}, \label{bz}
\end{align}
where $\eta_{x,z}$ and $\epsilon_{n_{x,z}}$ have the same definition as in Sec. II and $C_{E}^{(2D)}$ is defined in Eq. (\ref{ce2d}).
Furthermore, the integration in Eq. (\ref{blim_g2d}) diverges in the limit $z\rightarrow 0$, and we can remove this divergence via the similar approach as in Sec. III. To this end, we re-express Eq. (\ref{blim_g2d}) as
\begin{eqnarray} 
G_{0}^{\rm (2D)}(E ; \rho{\bf e}_z)
=-\int_{0}^{\infty} \frac{1}{4 \pi \beta} \cdot \exp\left(-\beta \kappa-\frac{1}{4\beta} \rho^{2}\right) d \beta-\int_{0}^{\infty}\left[\tilde{K}_{\rm 2D}(\rho;E,\beta)-Y(\rho;E,\beta)\right] d \beta
+Z(\rho;E), \label{blim_g2d}
\end{eqnarray}
with $\kappa$ being any positive number and 
\begin{eqnarray}
\tilde{K}_{\rm 2D}(\rho;E,\beta)=K_{\rm 2D}(\rho;E,\beta)-\frac{1}{4 \pi \beta} \cdot \exp\left(-\beta \kappa-\frac{1}{4\beta} \rho^{2}\right).
\end{eqnarray}
Furthermore, using the result 
\begin{eqnarray}
\int_{0}^{\infty} \frac{1}{4 \pi \beta} \cdot \exp\left(-\beta \kappa-\frac{1}{4\beta} \rho^{2}\right) d \beta
= -\frac{1}{2\pi} \ln \rho -\frac{1}{2\pi}\gamma-\frac{1}{4\pi} \ln \frac{\kappa}{4}+{\cal O}(\rho)\ \ \ ({\rm for}\ \rho>0),\label{xint}
\end{eqnarray}
with $\gamma=0.5772...$ being the Euler's constant,  we derive a result with the same form of Eq. (\ref{g2d}):
\begin{equation}
\lim_{\rho\rightarrow0}G_0^{\rm (2D)}(E;{\bf \bm{\rho}})=\frac{1}{2\pi}\ln\rho-2\left[W_{\rm 2D}(E)+\int_{0}^{+\infty}I_{\rm 2D}(E,\beta)d\beta\right].\label{g2d}
\end{equation}
In this step the functions $W_{\rm 2D}(E)$ and $I_{\rm 2D}(E,\beta)$ are given by
\begin{eqnarray}
W_{\rm 2D}(E)&=&-\frac{1}{4\pi}\gamma-\frac{1}{8\pi} \ln \frac{\kappa}{4}-\frac{1}{2}Z(0,E);\\
I_{\rm 2D}(E,\beta)&=&\frac{1}{2}\left[\tilde{K}_{\rm 2D}(0;E,\beta)-Y(0;E,\beta)\right].
\end{eqnarray}
Moreover,  with the method in Appendix A, we can derive the alternative expressions of $W_{\rm 2D}(E)$ and $I_{\rm 2D}(E,\beta)$, i.e., Eqs. (\ref{w2d}) and (\ref{iee2d}).


\section{Techniques for fast calculation of $J_{{\rm 2D}}(E)$ }

In this appendix we generalize the techniques shown in Sec.II.B and Appendix B to the 2D case.
We first  generalize Eqs. (\ref{dg22}-\ref{b23d}) to the 2D cases and prove Eq. (\ref{dg2d22}). This can be done via direct calculations with the method shown in Appendix B.  We separate the integration in Eq. (\ref{j2de}) into two parts, i.e.,
\begin{eqnarray}
\int_{0}^{\infty} I_{{\rm 2D}}(E,\beta)  d\beta=\int_{0}^{\Lambda} I_{{\rm 2D}}(E,\beta)  d\beta+\int_{\Lambda}^{\infty} I_{{\rm 2D}}(E,\beta)  d\beta, \label{bb12d}
\end{eqnarray}
with $\Lambda$ being an arbitrary positive number.
In addition, using the definition (\ref{iee2d}) of $I_{{\rm 2D}}(E,\beta)$, we immediately obtain
\begin{align}
&\int_{0}^{\Lambda} I_{{\rm 2D}}(E,\beta)  d\beta\nonumber\\
=&\int_{0}^{\Lambda}A_{{\rm 2D}}(E,\beta)d\beta
-\sum_{n_{x}\in C_{E}^{{\rm (2D)}}}
\int_{0}^{\Lambda}
 \frac{2^{n_{x}-\frac{5}{2}}\sqrt{\eta_x}e^{\beta(E-\epsilon_{n_{x}})}}{\Gamma\left(\frac{1-n_{x}}{2}\right)^{2}\Gamma(n_{x}+1)\sqrt{\sinh\beta}}
d\beta\nonumber\\
=& \int_{0}^{\Lambda}A_{{\rm 2D}}(E,\beta)d\beta+\sum_{(n_{x})\in C_{E}^{{\rm (2D)}}}
 \frac{2^{n_{x}-3}\sqrt{\eta_{x}}}
 {\Gamma\left(\frac{1-n_{x}}{2}\right)^{2}\Gamma(n_{x}+1)}
 \Bigg\{ -{\sqrt{\pi} \Gamma\left({1\over4}-{{E-\epsilon_{n_{x}}}\over2}\right)\over  \Gamma\left({3\over4}-{{E-\epsilon_{n_{x}}}\over2}\right)}+ \nonumber \\
& \frac{\Gamma\left({1\over4}-{{E-\epsilon_{n_{x}}}\over2}\right)}{\Gamma\left({5\over4}-{{E-\epsilon_{n_{x}}}\over2}\right)}e^{(E-\epsilon_{n_x}-{3\over2})\Lambda}\sqrt{e^{2\Lambda}-1}\times _{2}\!F^{1}\qty[1,\frac{3}{4}-\frac{E-\epsilon_{n_x}}{2},\frac{5}{4}-\frac{E-\epsilon_{n_x}}{2},
e^{-2\Lambda}]
 \Bigg\}
\label{bb22d}
\end{align}
with $A_{{\rm 2D}}(E,\beta)$ being defined in Eq. (\ref{a2d}).

Furthermore, using the method in Appendix B, we find that the function $I_{\rm 2D}(E,\beta)$ defined in Eq.~(\ref{iee2d}) has an alternative expression
\begin{eqnarray}
I_{\rm 2D}(E,\beta) & = & -\frac{1}{8\pi\beta}e^{-\kappa\beta}
+\sqrt{\frac{\eta_{x}}{4\sinh\beta}}\sum_{n_{x}\notin C_{E}^{\rm (2D)}}\left\{ \frac{2^{n_{x}-\frac{3}{2}}}{\Gamma\left(\frac{1-n_{x}}{2}\right)^{2}\Gamma(n_{x}+1)}e^{\beta(E-\epsilon_{n_{x}})}\right\}, \label{diee2d}
\end{eqnarray}
which is similar to Eq. (\ref{itd}).
Thus, doing the integration $\int_{\Lambda}^{\infty} I_{{\rm 2D}}(E,\beta)  d\beta$ in both sides of Eq. (\ref{diee2d}), we further obtain
\begin{eqnarray}
\int_{\Lambda}^{\infty} I_{{\rm 2D}}(E,\beta)  d\beta
&=&B^{(2)}_{\rm 2D}(E,\Lambda)-\frac{\Gamma(0,\kappa\Lambda_2)}{8\pi}
 \label{itdone2d}
\end{eqnarray}
where $\Gamma[a,z]$ is the incomplete Gamma function and $B^{(2)}_{\rm 2D}(E,\Lambda)$ is defined in Eq. (\ref{b2d2}).

As in Appendix B, substituting Eqs. (\ref{bb22d}, \ref{itdone2d}) into Eq. (\ref{bb12d}) and then into Eq. (\ref{j2de}), and further using Eqs. (\ref{w2d}), we can derive Eq. (\ref{dg2d22}).

In addition, the second technique  shown in Sec.II.B is the one based on Eqs. (\ref{dg2a}, \ref{dg22a}). It is clear that this technique can be directly generalized to the 2D case.

\end{widetext}



\begin{thebibliography}{53}%
\makeatletter
\providecommand \@ifxundefined [1]{%
 \@ifx{#1\undefined}
}%
\providecommand \@ifnum [1]{%
 \ifnum #1\expandafter \@firstoftwo
 \else \expandafter \@secondoftwo
 \fi
}%
\providecommand \@ifx [1]{%
 \ifx #1\expandafter \@firstoftwo
 \else \expandafter \@secondoftwo
 \fi
}%
\providecommand \natexlab [1]{#1}%
\providecommand \enquote  [1]{``#1''}%
\providecommand \bibnamefont  [1]{#1}%
\providecommand \bibfnamefont [1]{#1}%
\providecommand \citenamefont [1]{#1}%
\providecommand \href@noop [0]{\@secondoftwo}%
\providecommand \href [0]{\begingroup \@sanitize@url \@href}%
\providecommand \@href[1]{\@@startlink{#1}\@@href}%
\providecommand \@@href[1]{\endgroup#1\@@endlink}%
\providecommand \@sanitize@url [0]{\catcode `\\12\catcode `\$12\catcode
  `\&12\catcode `\#12\catcode `\^12\catcode `\_12\catcode `\%12\relax}%
\providecommand \@@startlink[1]{}%
\providecommand \@@endlink[0]{}%
\providecommand \url  [0]{\begingroup\@sanitize@url \@url }%
\providecommand \@url [1]{\endgroup\@href {#1}{\urlprefix }}%
\providecommand \urlprefix  [0]{URL }%
\providecommand \Eprint [0]{\href }%
\providecommand \doibase [0]{http://dx.doi.org/}%
\providecommand \selectlanguage [0]{\@gobble}%
\providecommand \bibinfo  [0]{\@secondoftwo}%
\providecommand \bibfield  [0]{\@secondoftwo}%
\providecommand \translation [1]{[#1]}%
\providecommand \BibitemOpen [0]{}%
\providecommand \bibitemStop [0]{}%
\providecommand \bibitemNoStop [0]{.\EOS\space}%
\providecommand \EOS [0]{\spacefactor3000\relax}%
\providecommand \BibitemShut  [1]{\csname bibitem#1\endcsname}%
\let\auto@bib@innerbib\@empty
\bibitem [{\citenamefont {Busch}\ \emph {et~al.}(1998)\citenamefont {Busch},
  \citenamefont {Englert}, \citenamefont {Rza{\.{z}}ewski},\ and\ \citenamefont
  {Wilkens}}]{busch}%
  \BibitemOpen
  \bibfield  {author} {\bibinfo {author} {\bibfnamefont {T.}~\bibnamefont
  {Busch}}, \bibinfo {author} {\bibfnamefont {B.-G.}\ \bibnamefont {Englert}},
  \bibinfo {author} {\bibfnamefont {K.}~\bibnamefont {Rza{\.{z}}ewski}}, \ and\
  \bibinfo {author} {\bibfnamefont {M.}~\bibnamefont {Wilkens}},\ }\href
  {\doibase 10.1023/A:1018705520999} {\bibfield  {journal} {\bibinfo  {journal}
  {Found. Phys.}\ }\textbf {\bibinfo {volume} {28}},\ \bibinfo {pages} {549}
  (\bibinfo {year} {1998})}\BibitemShut {NoStop}%
\bibitem [{\citenamefont {Idziaszek}\ and\ \citenamefont
  {Calarco}(2005)}]{calarco}%
  \BibitemOpen
  \bibfield  {author} {\bibinfo {author} {\bibfnamefont {Z.}~\bibnamefont
  {Idziaszek}}\ and\ \bibinfo {author} {\bibfnamefont {T.}~\bibnamefont
  {Calarco}},\ }\href {\doibase 10.1103/PhysRevA.71.050701} {\bibfield
  {journal} {\bibinfo  {journal} {Phys. Rev. A}\ }\textbf {\bibinfo {volume}
  {71}},\ \bibinfo {pages} {050701} (\bibinfo {year} {2005})}\BibitemShut
  {NoStop}%
\bibitem [{\citenamefont {Idziaszek}\ and\ \citenamefont
  {Calarco}(2006)}]{calarco2}%
  \BibitemOpen
  \bibfield  {author} {\bibinfo {author} {\bibfnamefont {Z.}~\bibnamefont
  {Idziaszek}}\ and\ \bibinfo {author} {\bibfnamefont {T.}~\bibnamefont
  {Calarco}},\ }\href {\doibase 10.1103/PhysRevA.74.022712} {\bibfield
  {journal} {\bibinfo  {journal} {Phys. Rev. A}\ }\textbf {\bibinfo {volume}
  {74}},\ \bibinfo {pages} {022712} (\bibinfo {year} {2006})}\BibitemShut
  {NoStop}%
\bibitem [{\citenamefont {Liang}\ and\ \citenamefont {Zhang}(2008)}]{liang}%
  \BibitemOpen
  \bibfield  {author} {\bibinfo {author} {\bibfnamefont {J.-J.}\ \bibnamefont
  {Liang}}\ and\ \bibinfo {author} {\bibfnamefont {C.}~\bibnamefont {Zhang}},\
  }\href {\doibase 10.1088/0031-8949/77/02/025302} {\bibfield  {journal}
  {\bibinfo  {journal} {Phys. Scr.}\ }\textbf {\bibinfo {volume} {77}},\
  \bibinfo {pages} {025302} (\bibinfo {year} {2008})}\BibitemShut {NoStop}%
\bibitem [{\citenamefont {Blume}(2012)}]{Blume_review}%
  \BibitemOpen
  \bibfield  {author} {\bibinfo {author} {\bibfnamefont {D.}~\bibnamefont
  {Blume}},\ }\href {\doibase 10.1088/0034-4885/75/4/046401} {\bibfield
  {journal} {\bibinfo  {journal} {Rep. Prog. Phys.}\ }\textbf {\bibinfo
  {volume} {75}},\ \bibinfo {pages} {046401} (\bibinfo {year}
  {2012})}\BibitemShut {NoStop}%
\bibitem [{\citenamefont {Grishkevich}\ and\ \citenamefont
  {Saenz}(2009)}]{szenz09}%
  \BibitemOpen
  \bibfield  {author} {\bibinfo {author} {\bibfnamefont {S.}~\bibnamefont
  {Grishkevich}}\ and\ \bibinfo {author} {\bibfnamefont {A.}~\bibnamefont
  {Saenz}},\ }\href {\doibase 10.1103/PhysRevA.80.013403} {\bibfield  {journal}
  {\bibinfo  {journal} {Phys. Rev. A}\ }\textbf {\bibinfo {volume} {80}},\
  \bibinfo {pages} {013403} (\bibinfo {year} {2009})}\BibitemShut {NoStop}%
\bibitem [{\citenamefont {Grishkevich}\ \emph {et~al.}(2011)\citenamefont
  {Grishkevich}, \citenamefont {Sala},\ and\ \citenamefont {Saenz}}]{szenz11}%
  \BibitemOpen
  \bibfield  {author} {\bibinfo {author} {\bibfnamefont {S.}~\bibnamefont
  {Grishkevich}}, \bibinfo {author} {\bibfnamefont {S.}~\bibnamefont {Sala}}, \
  and\ \bibinfo {author} {\bibfnamefont {A.}~\bibnamefont {Saenz}},\ }\href
  {\doibase 10.1103/PhysRevA.84.062710} {\bibfield  {journal} {\bibinfo
  {journal} {Phys. Rev. A}\ }\textbf {\bibinfo {volume} {84}},\ \bibinfo
  {pages} {062710} (\bibinfo {year} {2011})}\BibitemShut {NoStop}%
\bibitem [{\citenamefont {Sala}\ \emph {et~al.}(2013)\citenamefont {Sala},
  \citenamefont {Z\"urn}, \citenamefont {Lompe}, \citenamefont {Wenz},
  \citenamefont {Murmann}, \citenamefont {Serwane}, \citenamefont {Jochim},\
  and\ \citenamefont {Saenz}}]{szenz13}%
  \BibitemOpen
  \bibfield  {author} {\bibinfo {author} {\bibfnamefont {S.}~\bibnamefont
  {Sala}}, \bibinfo {author} {\bibfnamefont {G.}~\bibnamefont {Z\"urn}},
  \bibinfo {author} {\bibfnamefont {T.}~\bibnamefont {Lompe}}, \bibinfo
  {author} {\bibfnamefont {A.~N.}\ \bibnamefont {Wenz}}, \bibinfo {author}
  {\bibfnamefont {S.}~\bibnamefont {Murmann}}, \bibinfo {author} {\bibfnamefont
  {F.}~\bibnamefont {Serwane}}, \bibinfo {author} {\bibfnamefont
  {S.}~\bibnamefont {Jochim}}, \ and\ \bibinfo {author} {\bibfnamefont
  {A.}~\bibnamefont {Saenz}},\ }\href {\doibase 10.1103/PhysRevLett.110.203202}
  {\bibfield  {journal} {\bibinfo  {journal} {Phys. Rev. Lett.}\ }\textbf
  {\bibinfo {volume} {110}},\ \bibinfo {pages} {203202} (\bibinfo {year}
  {2013})}\BibitemShut {NoStop}%
\bibitem [{\citenamefont {Sala}\ and\ \citenamefont {Saenz}(2016)}]{szenz16}%
  \BibitemOpen
  \bibfield  {author} {\bibinfo {author} {\bibfnamefont {S.}~\bibnamefont
  {Sala}}\ and\ \bibinfo {author} {\bibfnamefont {A.}~\bibnamefont {Saenz}},\
  }\href {\doibase 10.1103/PhysRevA.94.022713} {\bibfield  {journal} {\bibinfo
  {journal} {Phys. Rev. A}\ }\textbf {\bibinfo {volume} {94}},\ \bibinfo
  {pages} {022713} (\bibinfo {year} {2016})}\BibitemShut {NoStop}%
\bibitem [{\citenamefont {Sun}\ \emph {et~al.}(2006)\citenamefont {Sun},
  \citenamefont {Zhang}, \citenamefont {Yi}, \citenamefont {Chapman},\ and\
  \citenamefont {You}}]{Liyou}%
  \BibitemOpen
  \bibfield  {author} {\bibinfo {author} {\bibfnamefont {B.}~\bibnamefont
  {Sun}}, \bibinfo {author} {\bibfnamefont {W.~X.}\ \bibnamefont {Zhang}},
  \bibinfo {author} {\bibfnamefont {S.}~\bibnamefont {Yi}}, \bibinfo {author}
  {\bibfnamefont {M.~S.}\ \bibnamefont {Chapman}}, \ and\ \bibinfo {author}
  {\bibfnamefont {L.}~\bibnamefont {You}},\ }\href {\doibase
  10.1103/PhysRevLett.97.123201} {\bibfield  {journal} {\bibinfo  {journal}
  {Phys. Rev. Lett.}\ }\textbf {\bibinfo {volume} {97}},\ \bibinfo {pages}
  {123201} (\bibinfo {year} {2006})}\BibitemShut {NoStop}%
\bibitem [{\citenamefont {Kehrberger}\ \emph {et~al.}(2018)\citenamefont
  {Kehrberger}, \citenamefont {Bolsinger},\ and\ \citenamefont
  {Schmelcher}}]{Schmelch1}%
  \BibitemOpen
  \bibfield  {author} {\bibinfo {author} {\bibfnamefont {L.~M.~A.}\
  \bibnamefont {Kehrberger}}, \bibinfo {author} {\bibfnamefont {V.~J.}\
  \bibnamefont {Bolsinger}}, \ and\ \bibinfo {author} {\bibfnamefont
  {P.}~\bibnamefont {Schmelcher}},\ }\href {\doibase
  10.1103/PhysRevA.97.013606} {\bibfield  {journal} {\bibinfo  {journal} {Phys.
  Rev. A}\ }\textbf {\bibinfo {volume} {97}},\ \bibinfo {pages} {013606}
  (\bibinfo {year} {2018})}\BibitemShut {NoStop}%
\bibitem [{\citenamefont {Bougas}\ \emph {et~al.}(2019)\citenamefont {Bougas},
  \citenamefont {Mistakidis},\ and\ \citenamefont {Schmelcher}}]{Schmelch2}%
  \BibitemOpen
  \bibfield  {author} {\bibinfo {author} {\bibfnamefont {G.}~\bibnamefont
  {Bougas}}, \bibinfo {author} {\bibfnamefont {S.~I.}\ \bibnamefont
  {Mistakidis}}, \ and\ \bibinfo {author} {\bibfnamefont {P.}~\bibnamefont
  {Schmelcher}},\ }\href {\doibase 10.1103/PhysRevA.100.053602} {\bibfield
  {journal} {\bibinfo  {journal} {Phys. Rev. A}\ }\textbf {\bibinfo {volume}
  {100}},\ \bibinfo {pages} {053602} (\bibinfo {year} {2019})}\BibitemShut
  {NoStop}%
\bibitem [{\citenamefont {Budewig}\ \emph {et~al.}(2019)\citenamefont
  {Budewig}, \citenamefont {Mistakidis},\ and\ \citenamefont
  {Schmelcher}}]{Schmelch3}%
  \BibitemOpen
  \bibfield  {author} {\bibinfo {author} {\bibfnamefont {L.}~\bibnamefont
  {Budewig}}, \bibinfo {author} {\bibfnamefont {S.~I.}\ \bibnamefont
  {Mistakidis}}, \ and\ \bibinfo {author} {\bibfnamefont {P.}~\bibnamefont
  {Schmelcher}},\ }\href {\doibase 10.1080/00268976.2019.1575995} {\bibfield
  {journal} {\bibinfo  {journal} {Molecular Physics}\ }\textbf {\bibinfo
  {volume} {117}},\ \bibinfo {pages} {2043} (\bibinfo {year} {2019})},\ \Eprint
  {http://arxiv.org/abs/https://doi.org/10.1080/00268976.2019.1575995}
  {https://doi.org/10.1080/00268976.2019.1575995} \BibitemShut {NoStop}%
\bibitem [{\citenamefont {Wenz}\ \emph {et~al.}(2013)\citenamefont {Wenz},
  \citenamefont {Z{\"u}rn}, \citenamefont {Murmann}, \citenamefont {Brouzos},
  \citenamefont {Lompe},\ and\ \citenamefont {Jochim}}]{Jochim}%
  \BibitemOpen
  \bibfield  {author} {\bibinfo {author} {\bibfnamefont {A.~N.}\ \bibnamefont
  {Wenz}}, \bibinfo {author} {\bibfnamefont {G.}~\bibnamefont {Z{\"u}rn}},
  \bibinfo {author} {\bibfnamefont {S.}~\bibnamefont {Murmann}}, \bibinfo
  {author} {\bibfnamefont {I.}~\bibnamefont {Brouzos}}, \bibinfo {author}
  {\bibfnamefont {T.}~\bibnamefont {Lompe}}, \ and\ \bibinfo {author}
  {\bibfnamefont {S.}~\bibnamefont {Jochim}},\ }\href {\doibase
  10.1126/science.1240516} {\bibfield  {journal} {\bibinfo  {journal}
  {Science}\ }\textbf {\bibinfo {volume} {342}},\ \bibinfo {pages} {457}
  (\bibinfo {year} {2013})}\BibitemShut {NoStop}%
\bibitem [{\citenamefont {Blume}\ and\ \citenamefont {Greene}(2002)}]{blume02}%
  \BibitemOpen
  \bibfield  {author} {\bibinfo {author} {\bibfnamefont {D.}~\bibnamefont
  {Blume}}\ and\ \bibinfo {author} {\bibfnamefont {C.~H.}\ \bibnamefont
  {Greene}},\ }\href {\doibase 10.1103/PhysRevA.66.013601} {\bibfield
  {journal} {\bibinfo  {journal} {Phys. Rev. A}\ }\textbf {\bibinfo {volume}
  {66}},\ \bibinfo {pages} {013601} (\bibinfo {year} {2002})}\BibitemShut
  {NoStop}%
\bibitem [{\citenamefont {Liu}\ \emph {et~al.}(2009)\citenamefont {Liu},
  \citenamefont {Hu},\ and\ \citenamefont {Drummond}}]{HuiHu_virial}%
  \BibitemOpen
  \bibfield  {author} {\bibinfo {author} {\bibfnamefont {X.-J.}\ \bibnamefont
  {Liu}}, \bibinfo {author} {\bibfnamefont {H.}~\bibnamefont {Hu}}, \ and\
  \bibinfo {author} {\bibfnamefont {P.~D.}\ \bibnamefont {Drummond}},\ }\href
  {\doibase 10.1103/PhysRevLett.102.160401} {\bibfield  {journal} {\bibinfo
  {journal} {Phys. Rev. Lett.}\ }\textbf {\bibinfo {volume} {102}},\ \bibinfo
  {pages} {160401} (\bibinfo {year} {2009})}\BibitemShut {NoStop}%
\bibitem [{\citenamefont {von Stecher}\ \emph {et~al.}(2008)\citenamefont {von
  Stecher}, \citenamefont {Greene},\ and\ \citenamefont {Blume}}]{blume08}%
  \BibitemOpen
  \bibfield  {author} {\bibinfo {author} {\bibfnamefont {J.}~\bibnamefont {von
  Stecher}}, \bibinfo {author} {\bibfnamefont {C.~H.}\ \bibnamefont {Greene}},
  \ and\ \bibinfo {author} {\bibfnamefont {D.}~\bibnamefont {Blume}},\ }\href
  {\doibase 10.1103/PhysRevA.77.043619} {\bibfield  {journal} {\bibinfo
  {journal} {Phys. Rev. A}\ }\textbf {\bibinfo {volume} {77}},\ \bibinfo
  {pages} {043619} (\bibinfo {year} {2008})}\BibitemShut {NoStop}%
\bibitem [{\citenamefont {Daily}\ and\ \citenamefont {Blume}(2010)}]{blume10}%
  \BibitemOpen
  \bibfield  {author} {\bibinfo {author} {\bibfnamefont {K.~M.}\ \bibnamefont
  {Daily}}\ and\ \bibinfo {author} {\bibfnamefont {D.}~\bibnamefont {Blume}},\
  }\href {\doibase 10.1103/PhysRevA.81.053615} {\bibfield  {journal} {\bibinfo
  {journal} {Phys. Rev. A}\ }\textbf {\bibinfo {volume} {81}},\ \bibinfo
  {pages} {053615} (\bibinfo {year} {2010})}\BibitemShut {NoStop}%
\bibitem [{\citenamefont {Gharashi}\ \emph {et~al.}(2012)\citenamefont
  {Gharashi}, \citenamefont {Daily},\ and\ \citenamefont {Blume}}]{blume12}%
  \BibitemOpen
  \bibfield  {author} {\bibinfo {author} {\bibfnamefont {S.~E.}\ \bibnamefont
  {Gharashi}}, \bibinfo {author} {\bibfnamefont {K.~M.}\ \bibnamefont {Daily}},
  \ and\ \bibinfo {author} {\bibfnamefont {D.}~\bibnamefont {Blume}},\ }\href
  {\doibase 10.1103/PhysRevA.86.042702} {\bibfield  {journal} {\bibinfo
  {journal} {Phys. Rev. A}\ }\textbf {\bibinfo {volume} {86}},\ \bibinfo
  {pages} {042702} (\bibinfo {year} {2012})}\BibitemShut {NoStop}%
\bibitem [{\citenamefont {Liu}(2013)}]{xiaji-virial}%
  \BibitemOpen
  \bibfield  {author} {\bibinfo {author} {\bibfnamefont {X.-J.}\ \bibnamefont
  {Liu}},\ }\href {\doibase https://doi.org/10.1016/j.physrep.2012.10.004}
  {\bibfield  {journal} {\bibinfo  {journal} {Phys. Rep.}\ }\textbf {\bibinfo
  {volume} {524}},\ \bibinfo {pages} {37 } (\bibinfo {year}
  {2013})}\BibitemShut {NoStop}%
\bibitem [{\citenamefont {Yin}\ \emph {et~al.}(2014{\natexlab{a}})\citenamefont
  {Yin}, \citenamefont {Blume}, \citenamefont {Johnson},\ and\ \citenamefont
  {Tiesinga}}]{blume14}%
  \BibitemOpen
  \bibfield  {author} {\bibinfo {author} {\bibfnamefont {X.~Y.}\ \bibnamefont
  {Yin}}, \bibinfo {author} {\bibfnamefont {D.}~\bibnamefont {Blume}}, \bibinfo
  {author} {\bibfnamefont {P.~R.}\ \bibnamefont {Johnson}}, \ and\ \bibinfo
  {author} {\bibfnamefont {E.}~\bibnamefont {Tiesinga}},\ }\href {\doibase
  10.1103/PhysRevA.90.043631} {\bibfield  {journal} {\bibinfo  {journal} {Phys.
  Rev. A}\ }\textbf {\bibinfo {volume} {90}},\ \bibinfo {pages} {043631}
  (\bibinfo {year} {2014}{\natexlab{a}})}\BibitemShut {NoStop}%
\bibitem [{\citenamefont {Yin}\ \emph {et~al.}(2014{\natexlab{b}})\citenamefont
  {Yin}, \citenamefont {Blume}, \citenamefont {Johnson},\ and\ \citenamefont
  {Tiesinga}}]{XYYin14}%
  \BibitemOpen
  \bibfield  {author} {\bibinfo {author} {\bibfnamefont {X.~Y.}\ \bibnamefont
  {Yin}}, \bibinfo {author} {\bibfnamefont {D.}~\bibnamefont {Blume}}, \bibinfo
  {author} {\bibfnamefont {P.~R.}\ \bibnamefont {Johnson}}, \ and\ \bibinfo
  {author} {\bibfnamefont {E.}~\bibnamefont {Tiesinga}},\ }\href {\doibase
  10.1103/PhysRevA.90.043631} {\bibfield  {journal} {\bibinfo  {journal} {Phys.
  Rev. A}\ }\textbf {\bibinfo {volume} {90}},\ \bibinfo {pages} {043631}
  (\bibinfo {year} {2014}{\natexlab{b}})}\BibitemShut {NoStop}%
\bibitem [{\citenamefont {Peng}\ \emph {et~al.}(2014)\citenamefont {Peng},
  \citenamefont {Zhao},\ and\ \citenamefont {Jiang}}]{shiguo-virial}%
  \BibitemOpen
  \bibfield  {author} {\bibinfo {author} {\bibfnamefont {S.-G.}\ \bibnamefont
  {Peng}}, \bibinfo {author} {\bibfnamefont {S.-H.}\ \bibnamefont {Zhao}}, \
  and\ \bibinfo {author} {\bibfnamefont {K.}~\bibnamefont {Jiang}},\ }\href
  {\doibase 10.1103/PhysRevA.89.013603} {\bibfield  {journal} {\bibinfo
  {journal} {Phys. Rev. A}\ }\textbf {\bibinfo {volume} {89}},\ \bibinfo
  {pages} {013603} (\bibinfo {year} {2014})}\BibitemShut {NoStop}%
\bibitem [{\citenamefont {Yin}\ and\ \citenamefont {Blume}(2015)}]{blume15}%
  \BibitemOpen
  \bibfield  {author} {\bibinfo {author} {\bibfnamefont {X.~Y.}\ \bibnamefont
  {Yin}}\ and\ \bibinfo {author} {\bibfnamefont {D.}~\bibnamefont {Blume}},\
  }\href {\doibase 10.1103/PhysRevA.92.013608} {\bibfield  {journal} {\bibinfo
  {journal} {Phys. Rev. A}\ }\textbf {\bibinfo {volume} {92}},\ \bibinfo
  {pages} {013608} (\bibinfo {year} {2015})}\BibitemShut {NoStop}%
\bibitem [{\citenamefont {Blume}\ \emph {et~al.}(2018)\citenamefont {Blume},
  \citenamefont {Sze},\ and\ \citenamefont {Bohn}}]{blume18}%
  \BibitemOpen
  \bibfield  {author} {\bibinfo {author} {\bibfnamefont {D.}~\bibnamefont
  {Blume}}, \bibinfo {author} {\bibfnamefont {M.~W.~C.}\ \bibnamefont {Sze}}, \
  and\ \bibinfo {author} {\bibfnamefont {J.~L.}\ \bibnamefont {Bohn}},\ }\href
  {\doibase 10.1103/PhysRevA.97.033621} {\bibfield  {journal} {\bibinfo
  {journal} {Phys. Rev. A}\ }\textbf {\bibinfo {volume} {97}},\ \bibinfo
  {pages} {033621} (\bibinfo {year} {2018})}\BibitemShut {NoStop}%
\bibitem [{\citenamefont {Yin}\ \emph {et~al.}(2020)\citenamefont {Yin},
  \citenamefont {Hu},\ and\ \citenamefont {Liu}}]{XYYin20}%
  \BibitemOpen
  \bibfield  {author} {\bibinfo {author} {\bibfnamefont {X.~Y.}\ \bibnamefont
  {Yin}}, \bibinfo {author} {\bibfnamefont {H.}~\bibnamefont {Hu}}, \ and\
  \bibinfo {author} {\bibfnamefont {X.-J.}\ \bibnamefont {Liu}},\ }\href
  {\doibase 10.1103/PhysRevLett.124.013401} {\bibfield  {journal} {\bibinfo
  {journal} {Phys. Rev. Lett.}\ }\textbf {\bibinfo {volume} {124}},\ \bibinfo
  {pages} {013401} (\bibinfo {year} {2020})}\BibitemShut {NoStop}%
\bibitem [{\citenamefont {Scazza}\ \emph {et~al.}(2014)\citenamefont {Scazza},
  \citenamefont {Hofrichter}, \citenamefont {H{\"{o}}fer}, \citenamefont {{De
  Groot}}, \citenamefont {Bloch},\ and\ \citenamefont
  {F{\"{o}}lling}}]{Scazza2014}%
  \BibitemOpen
  \bibfield  {author} {\bibinfo {author} {\bibfnamefont {F.}~\bibnamefont
  {Scazza}}, \bibinfo {author} {\bibfnamefont {C.}~\bibnamefont {Hofrichter}},
  \bibinfo {author} {\bibfnamefont {M.}~\bibnamefont {H{\"{o}}fer}}, \bibinfo
  {author} {\bibfnamefont {P.~C.}\ \bibnamefont {{De Groot}}}, \bibinfo
  {author} {\bibfnamefont {I.}~\bibnamefont {Bloch}}, \ and\ \bibinfo {author}
  {\bibfnamefont {S.}~\bibnamefont {F{\"{o}}lling}},\ }\href
  {https://www.nature.com/articles/nphys3061} {\bibfield  {journal} {\bibinfo
  {journal} {Nat. Phys.}\ }\textbf {\bibinfo {volume} {10}},\ \bibinfo {pages}
  {779} (\bibinfo {year} {2014})}\BibitemShut {NoStop}%
\bibitem [{\citenamefont {Cappellini}\ \emph {et~al.}(2014)\citenamefont
  {Cappellini}, \citenamefont {Mancini}, \citenamefont {Pagano}, \citenamefont
  {Lombardi}, \citenamefont {Livi}, \citenamefont {Siciliani~de Cumis},
  \citenamefont {Cancio}, \citenamefont {Pizzocaro}, \citenamefont {Calonico},
  \citenamefont {Levi}, \citenamefont {Sias}, \citenamefont {Catani},
  \citenamefont {Inguscio},\ and\ \citenamefont {Fallani}}]{Cappellini2014}%
  \BibitemOpen
  \bibfield  {author} {\bibinfo {author} {\bibfnamefont {G.}~\bibnamefont
  {Cappellini}}, \bibinfo {author} {\bibfnamefont {M.}~\bibnamefont {Mancini}},
  \bibinfo {author} {\bibfnamefont {G.}~\bibnamefont {Pagano}}, \bibinfo
  {author} {\bibfnamefont {P.}~\bibnamefont {Lombardi}}, \bibinfo {author}
  {\bibfnamefont {L.}~\bibnamefont {Livi}}, \bibinfo {author} {\bibfnamefont
  {M.}~\bibnamefont {Siciliani~de Cumis}}, \bibinfo {author} {\bibfnamefont
  {P.}~\bibnamefont {Cancio}}, \bibinfo {author} {\bibfnamefont
  {M.}~\bibnamefont {Pizzocaro}}, \bibinfo {author} {\bibfnamefont
  {D.}~\bibnamefont {Calonico}}, \bibinfo {author} {\bibfnamefont
  {F.}~\bibnamefont {Levi}}, \bibinfo {author} {\bibfnamefont {C.}~\bibnamefont
  {Sias}}, \bibinfo {author} {\bibfnamefont {J.}~\bibnamefont {Catani}},
  \bibinfo {author} {\bibfnamefont {M.}~\bibnamefont {Inguscio}}, \ and\
  \bibinfo {author} {\bibfnamefont {L.}~\bibnamefont {Fallani}},\ }\href
  {\doibase 10.1103/PhysRevLett.113.120402} {\bibfield  {journal} {\bibinfo
  {journal} {Phys. Rev. Lett.}\ }\textbf {\bibinfo {volume} {113}},\ \bibinfo
  {pages} {120402} (\bibinfo {year} {2014})}\BibitemShut {NoStop}%
\bibitem [{\citenamefont {Norcia}\ \emph {et~al.}(2018)\citenamefont {Norcia},
  \citenamefont {Young},\ and\ \citenamefont {Kaufman}}]{Norcia18}%
  \BibitemOpen
  \bibfield  {author} {\bibinfo {author} {\bibfnamefont {M.~A.}\ \bibnamefont
  {Norcia}}, \bibinfo {author} {\bibfnamefont {A.~W.}\ \bibnamefont {Young}}, \
  and\ \bibinfo {author} {\bibfnamefont {A.~M.}\ \bibnamefont {Kaufman}},\
  }\href {\doibase 10.1103/PhysRevX.8.041054} {\bibfield  {journal} {\bibinfo
  {journal} {Phys. Rev. X}\ }\textbf {\bibinfo {volume} {8}},\ \bibinfo {pages}
  {041054} (\bibinfo {year} {2018})}\BibitemShut {NoStop}%
\bibitem [{\citenamefont {Cooper}\ \emph {et~al.}(2018)\citenamefont {Cooper},
  \citenamefont {Covey}, \citenamefont {Madjarov}, \citenamefont {Porsev},
  \citenamefont {Safronova},\ and\ \citenamefont {Endres}}]{Cooper18}%
  \BibitemOpen
  \bibfield  {author} {\bibinfo {author} {\bibfnamefont {A.}~\bibnamefont
  {Cooper}}, \bibinfo {author} {\bibfnamefont {J.~P.}\ \bibnamefont {Covey}},
  \bibinfo {author} {\bibfnamefont {I.~S.}\ \bibnamefont {Madjarov}}, \bibinfo
  {author} {\bibfnamefont {S.~G.}\ \bibnamefont {Porsev}}, \bibinfo {author}
  {\bibfnamefont {M.~S.}\ \bibnamefont {Safronova}}, \ and\ \bibinfo {author}
  {\bibfnamefont {M.}~\bibnamefont {Endres}},\ }\href {\doibase
  10.1103/PhysRevX.8.041055} {\bibfield  {journal} {\bibinfo  {journal} {Phys.
  Rev. X}\ }\textbf {\bibinfo {volume} {8}},\ \bibinfo {pages} {041055}
  (\bibinfo {year} {2018})}\BibitemShut {NoStop}%
\bibitem [{\citenamefont {Cappellini}\ \emph {et~al.}(2019)\citenamefont
  {Cappellini}, \citenamefont {Livi}, \citenamefont {Franchi}, \citenamefont
  {Tusi}, \citenamefont {Benedicto~Orenes}, \citenamefont {Inguscio},
  \citenamefont {Catani},\ and\ \citenamefont {Fallani}}]{Cappellini2019}%
  \BibitemOpen
  \bibfield  {author} {\bibinfo {author} {\bibfnamefont {G.}~\bibnamefont
  {Cappellini}}, \bibinfo {author} {\bibfnamefont {L.~F.}\ \bibnamefont
  {Livi}}, \bibinfo {author} {\bibfnamefont {L.}~\bibnamefont {Franchi}},
  \bibinfo {author} {\bibfnamefont {D.}~\bibnamefont {Tusi}}, \bibinfo {author}
  {\bibfnamefont {D.}~\bibnamefont {Benedicto~Orenes}}, \bibinfo {author}
  {\bibfnamefont {M.}~\bibnamefont {Inguscio}}, \bibinfo {author}
  {\bibfnamefont {J.}~\bibnamefont {Catani}}, \ and\ \bibinfo {author}
  {\bibfnamefont {L.}~\bibnamefont {Fallani}},\ }\href {\doibase
  10.1103/PhysRevX.9.011028} {\bibfield  {journal} {\bibinfo  {journal} {Phys.
  Rev. X}\ }\textbf {\bibinfo {volume} {9}},\ \bibinfo {pages} {011028}
  (\bibinfo {year} {2019})}\BibitemShut {NoStop}%
\bibitem [{\citenamefont {Guan}\ \emph {et~al.}(2019)\citenamefont {Guan},
  \citenamefont {Klinkhamer}, \citenamefont {Klemt}, \citenamefont {Becher},
  \citenamefont {Bergschneider}, \citenamefont {Preiss}, \citenamefont
  {Jochim},\ and\ \citenamefont {Blume}}]{guan2b}%
  \BibitemOpen
  \bibfield  {author} {\bibinfo {author} {\bibfnamefont {Q.}~\bibnamefont
  {Guan}}, \bibinfo {author} {\bibfnamefont {V.}~\bibnamefont {Klinkhamer}},
  \bibinfo {author} {\bibfnamefont {R.}~\bibnamefont {Klemt}}, \bibinfo
  {author} {\bibfnamefont {J.~H.}\ \bibnamefont {Becher}}, \bibinfo {author}
  {\bibfnamefont {A.}~\bibnamefont {Bergschneider}}, \bibinfo {author}
  {\bibfnamefont {P.~M.}\ \bibnamefont {Preiss}}, \bibinfo {author}
  {\bibfnamefont {S.}~\bibnamefont {Jochim}}, \ and\ \bibinfo {author}
  {\bibfnamefont {D.}~\bibnamefont {Blume}},\ }\href {\doibase
  10.1103/PhysRevLett.122.083401} {\bibfield  {journal} {\bibinfo  {journal}
  {Phys. Rev. Lett.}\ }\textbf {\bibinfo {volume} {122}},\ \bibinfo {pages}
  {083401} (\bibinfo {year} {2019})}\BibitemShut {NoStop}%
\bibitem [{\citenamefont {Liu}\ \emph {et~al.}(2018)\citenamefont {Liu},
  \citenamefont {Hood}, \citenamefont {Yu}, \citenamefont {Zhang},
  \citenamefont {Hutzler}, \citenamefont {Rosenband},\ and\ \citenamefont
  {Ni}}]{Ni1}%
  \BibitemOpen
  \bibfield  {author} {\bibinfo {author} {\bibfnamefont {L.~R.}\ \bibnamefont
  {Liu}}, \bibinfo {author} {\bibfnamefont {J.~D.}\ \bibnamefont {Hood}},
  \bibinfo {author} {\bibfnamefont {Y.}~\bibnamefont {Yu}}, \bibinfo {author}
  {\bibfnamefont {J.~T.}\ \bibnamefont {Zhang}}, \bibinfo {author}
  {\bibfnamefont {N.~R.}\ \bibnamefont {Hutzler}}, \bibinfo {author}
  {\bibfnamefont {T.}~\bibnamefont {Rosenband}}, \ and\ \bibinfo {author}
  {\bibfnamefont {K.-K.}\ \bibnamefont {Ni}},\ }\href {\doibase
  10.1126/science.aar7797} {\bibfield  {journal} {\bibinfo  {journal}
  {Science}\ }\textbf {\bibinfo {volume} {360}},\ \bibinfo {pages} {900}
  (\bibinfo {year} {2018})}\BibitemShut {NoStop}%
\bibitem [{\citenamefont {Anderegg}\ \emph {et~al.}(2019)\citenamefont
  {Anderegg}, \citenamefont {Cheuk}, \citenamefont {Bao}, \citenamefont
  {Burchesky}, \citenamefont {Ketterle}, \citenamefont {Ni},\ and\
  \citenamefont {Doyle}}]{Ni2}%
  \BibitemOpen
  \bibfield  {author} {\bibinfo {author} {\bibfnamefont {L.}~\bibnamefont
  {Anderegg}}, \bibinfo {author} {\bibfnamefont {L.~W.}\ \bibnamefont {Cheuk}},
  \bibinfo {author} {\bibfnamefont {Y.}~\bibnamefont {Bao}}, \bibinfo {author}
  {\bibfnamefont {S.}~\bibnamefont {Burchesky}}, \bibinfo {author}
  {\bibfnamefont {W.}~\bibnamefont {Ketterle}}, \bibinfo {author}
  {\bibfnamefont {K.-K.}\ \bibnamefont {Ni}}, \ and\ \bibinfo {author}
  {\bibfnamefont {J.~M.}\ \bibnamefont {Doyle}},\ }\href {\doibase
  10.1126/science.aax1265} {\bibfield  {journal} {\bibinfo  {journal}
  {Science}\ }\textbf {\bibinfo {volume} {365}},\ \bibinfo {pages} {1156}
  (\bibinfo {year} {2019})}\BibitemShut {NoStop}%
\bibitem [{\citenamefont {Hood}\ \emph {et~al.}(2019)\citenamefont {Hood},
  \citenamefont {Yu}, \citenamefont {Lin}, \citenamefont {Zhang}, \citenamefont
  {Wang}, \citenamefont {Liu}, \citenamefont {Gao},\ and\ \citenamefont
  {Ni}}]{Ni3}%
  \BibitemOpen
  \bibfield  {author} {\bibinfo {author} {\bibfnamefont {J.~D.}\ \bibnamefont
  {Hood}}, \bibinfo {author} {\bibfnamefont {Y.}~\bibnamefont {Yu}}, \bibinfo
  {author} {\bibfnamefont {Y.-W.}\ \bibnamefont {Lin}}, \bibinfo {author}
  {\bibfnamefont {J.~T.}\ \bibnamefont {Zhang}}, \bibinfo {author}
  {\bibfnamefont {K.}~\bibnamefont {Wang}}, \bibinfo {author} {\bibfnamefont
  {L.~R.}\ \bibnamefont {Liu}}, \bibinfo {author} {\bibfnamefont
  {B.}~\bibnamefont {Gao}}, \ and\ \bibinfo {author} {\bibfnamefont {K.-K.}\
  \bibnamefont {Ni}},\ }\href@noop {} {\enquote {\bibinfo {title} {Multichannel
  interactions of two atoms in an optical tweezer},}\ } (\bibinfo {year}
  {2019}),\ \Eprint {http://arxiv.org/abs/1907.11226} {arXiv:1907.11226}
  \BibitemShut {NoStop}%
\bibitem [{\citenamefont {Wang}\ \emph {et~al.}(2019)\citenamefont {Wang},
  \citenamefont {He}, \citenamefont {Guo}, \citenamefont {Xu}, \citenamefont
  {Sheng}, \citenamefont {Zhuang}, \citenamefont {Xiong}, \citenamefont {Liu},
  \citenamefont {Wang},\ and\ \citenamefont {Zhan}}]{XDHe}%
  \BibitemOpen
  \bibfield  {author} {\bibinfo {author} {\bibfnamefont {K.}~\bibnamefont
  {Wang}}, \bibinfo {author} {\bibfnamefont {X.}~\bibnamefont {He}}, \bibinfo
  {author} {\bibfnamefont {R.}~\bibnamefont {Guo}}, \bibinfo {author}
  {\bibfnamefont {P.}~\bibnamefont {Xu}}, \bibinfo {author} {\bibfnamefont
  {C.}~\bibnamefont {Sheng}}, \bibinfo {author} {\bibfnamefont
  {J.}~\bibnamefont {Zhuang}}, \bibinfo {author} {\bibfnamefont
  {Z.}~\bibnamefont {Xiong}}, \bibinfo {author} {\bibfnamefont
  {M.}~\bibnamefont {Liu}}, \bibinfo {author} {\bibfnamefont {J.}~\bibnamefont
  {Wang}}, \ and\ \bibinfo {author} {\bibfnamefont {M.}~\bibnamefont {Zhan}},\
  }\href {\doibase 10.1103/PhysRevA.100.063429} {\bibfield  {journal} {\bibinfo
   {journal} {Phys. Rev. A}\ }\textbf {\bibinfo {volume} {100}},\ \bibinfo
  {pages} {063429} (\bibinfo {year} {2019})}\BibitemShut {NoStop}%
\bibitem [{\citenamefont {Chang}\ \emph {et~al.}(2018)\citenamefont {Chang},
  \citenamefont {Douglas}, \citenamefont {Gonz\'alez-Tudela}, \citenamefont
  {Hung},\ and\ \citenamefont {Kimble}}]{nano1}%
  \BibitemOpen
  \bibfield  {author} {\bibinfo {author} {\bibfnamefont {D.~E.}\ \bibnamefont
  {Chang}}, \bibinfo {author} {\bibfnamefont {J.~S.}\ \bibnamefont {Douglas}},
  \bibinfo {author} {\bibfnamefont {A.}~\bibnamefont {Gonz\'alez-Tudela}},
  \bibinfo {author} {\bibfnamefont {C.-L.}\ \bibnamefont {Hung}}, \ and\
  \bibinfo {author} {\bibfnamefont {H.~J.}\ \bibnamefont {Kimble}},\ }\href
  {\doibase 10.1103/RevModPhys.90.031002} {\bibfield  {journal} {\bibinfo
  {journal} {Rev. Mod. Phys.}\ }\textbf {\bibinfo {volume} {90}},\ \bibinfo
  {pages} {031002} (\bibinfo {year} {2018})}\BibitemShut {NoStop}%
\bibitem [{\citenamefont {Meng}\ \emph {et~al.}(2018)\citenamefont {Meng},
  \citenamefont {Dareau}, \citenamefont {Schneeweiss},\ and\ \citenamefont
  {Rauschenbeutel}}]{nano2}%
  \BibitemOpen
  \bibfield  {author} {\bibinfo {author} {\bibfnamefont {Y.}~\bibnamefont
  {Meng}}, \bibinfo {author} {\bibfnamefont {A.}~\bibnamefont {Dareau}},
  \bibinfo {author} {\bibfnamefont {P.}~\bibnamefont {Schneeweiss}}, \ and\
  \bibinfo {author} {\bibfnamefont {A.}~\bibnamefont {Rauschenbeutel}},\ }\href
  {\doibase 10.1103/PhysRevX.8.031054} {\bibfield  {journal} {\bibinfo
  {journal} {Phys. Rev. X}\ }\textbf {\bibinfo {volume} {8}},\ \bibinfo {pages}
  {031054} (\bibinfo {year} {2018})}\BibitemShut {NoStop}%
\bibitem [{\citenamefont {Dareau}\ \emph {et~al.}(2018)\citenamefont {Dareau},
  \citenamefont {Meng}, \citenamefont {Schneeweiss},\ and\ \citenamefont
  {Rauschenbeutel}}]{nano3}%
  \BibitemOpen
  \bibfield  {author} {\bibinfo {author} {\bibfnamefont {A.}~\bibnamefont
  {Dareau}}, \bibinfo {author} {\bibfnamefont {Y.}~\bibnamefont {Meng}},
  \bibinfo {author} {\bibfnamefont {P.}~\bibnamefont {Schneeweiss}}, \ and\
  \bibinfo {author} {\bibfnamefont {A.}~\bibnamefont {Rauschenbeutel}},\ }\href
  {\doibase 10.1103/PhysRevLett.121.253603} {\bibfield  {journal} {\bibinfo
  {journal} {Phys. Rev. Lett.}\ }\textbf {\bibinfo {volume} {121}},\ \bibinfo
  {pages} {253603} (\bibinfo {year} {2018})}\BibitemShut {NoStop}%
\bibitem [{\citenamefont {Ospelkaus}\ \emph {et~al.}(2006)\citenamefont
  {Ospelkaus}, \citenamefont {Ospelkaus}, \citenamefont {Humbert},
  \citenamefont {Ernst}, \citenamefont {Sengstock},\ and\ \citenamefont
  {Bongs}}]{Ospelkaus06}%
  \BibitemOpen
  \bibfield  {author} {\bibinfo {author} {\bibfnamefont {C.}~\bibnamefont
  {Ospelkaus}}, \bibinfo {author} {\bibfnamefont {S.}~\bibnamefont
  {Ospelkaus}}, \bibinfo {author} {\bibfnamefont {L.}~\bibnamefont {Humbert}},
  \bibinfo {author} {\bibfnamefont {P.}~\bibnamefont {Ernst}}, \bibinfo
  {author} {\bibfnamefont {K.}~\bibnamefont {Sengstock}}, \ and\ \bibinfo
  {author} {\bibfnamefont {K.}~\bibnamefont {Bongs}},\ }\href {\doibase
  10.1103/PhysRevLett.97.120402} {\bibfield  {journal} {\bibinfo  {journal}
  {Phys. Rev. Lett.}\ }\textbf {\bibinfo {volume} {97}},\ \bibinfo {pages}
  {120402} (\bibinfo {year} {2006})}\BibitemShut {NoStop}%
\bibitem [{\citenamefont {St\"oferle}\ \emph {et~al.}(2006)\citenamefont
  {St\"oferle}, \citenamefont {Moritz}, \citenamefont {G\"unter}, \citenamefont
  {K\"ohl},\ and\ \citenamefont {Esslinger}}]{Stoferle06}%
  \BibitemOpen
  \bibfield  {author} {\bibinfo {author} {\bibfnamefont {T.}~\bibnamefont
  {St\"oferle}}, \bibinfo {author} {\bibfnamefont {H.}~\bibnamefont {Moritz}},
  \bibinfo {author} {\bibfnamefont {K.}~\bibnamefont {G\"unter}}, \bibinfo
  {author} {\bibfnamefont {M.}~\bibnamefont {K\"ohl}}, \ and\ \bibinfo {author}
  {\bibfnamefont {T.}~\bibnamefont {Esslinger}},\ }\href {\doibase
  10.1103/PhysRevLett.96.030401} {\bibfield  {journal} {\bibinfo  {journal}
  {Phys. Rev. Lett.}\ }\textbf {\bibinfo {volume} {96}},\ \bibinfo {pages}
  {030401} (\bibinfo {year} {2006})}\BibitemShut {NoStop}%
\bibitem [{\citenamefont {Mark}\ \emph {et~al.}(2011)\citenamefont {Mark},
  \citenamefont {Haller}, \citenamefont {Lauber}, \citenamefont {Danzl},
  \citenamefont {Daley},\ and\ \citenamefont {N\"agerl}}]{Mark11}%
  \BibitemOpen
  \bibfield  {author} {\bibinfo {author} {\bibfnamefont {M.~J.}\ \bibnamefont
  {Mark}}, \bibinfo {author} {\bibfnamefont {E.}~\bibnamefont {Haller}},
  \bibinfo {author} {\bibfnamefont {K.}~\bibnamefont {Lauber}}, \bibinfo
  {author} {\bibfnamefont {J.~G.}\ \bibnamefont {Danzl}}, \bibinfo {author}
  {\bibfnamefont {A.~J.}\ \bibnamefont {Daley}}, \ and\ \bibinfo {author}
  {\bibfnamefont {H.-C.}\ \bibnamefont {N\"agerl}},\ }\href {\doibase
  10.1103/PhysRevLett.107.175301} {\bibfield  {journal} {\bibinfo  {journal}
  {Phys. Rev. Lett.}\ }\textbf {\bibinfo {volume} {107}},\ \bibinfo {pages}
  {175301} (\bibinfo {year} {2011})}\BibitemShut {NoStop}%
\bibitem [{\citenamefont {Riegger}\ \emph {et~al.}(2018)\citenamefont
  {Riegger}, \citenamefont {Darkwah~Oppong}, \citenamefont {H\"ofer},
  \citenamefont {Fernandes}, \citenamefont {Bloch},\ and\ \citenamefont
  {F\"olling}}]{Riegger18}%
  \BibitemOpen
  \bibfield  {author} {\bibinfo {author} {\bibfnamefont {L.}~\bibnamefont
  {Riegger}}, \bibinfo {author} {\bibfnamefont {N.}~\bibnamefont
  {Darkwah~Oppong}}, \bibinfo {author} {\bibfnamefont {M.}~\bibnamefont
  {H\"ofer}}, \bibinfo {author} {\bibfnamefont {D.~R.}\ \bibnamefont
  {Fernandes}}, \bibinfo {author} {\bibfnamefont {I.}~\bibnamefont {Bloch}}, \
  and\ \bibinfo {author} {\bibfnamefont {S.}~\bibnamefont {F\"olling}},\ }\href
  {\doibase 10.1103/PhysRevLett.120.143601} {\bibfield  {journal} {\bibinfo
  {journal} {Phys. Rev. Lett.}\ }\textbf {\bibinfo {volume} {120}},\ \bibinfo
  {pages} {143601} (\bibinfo {year} {2018})}\BibitemShut {NoStop}%
\bibitem [{\citenamefont {Chapurin}\ \emph {et~al.}(2019)\citenamefont
  {Chapurin}, \citenamefont {Xie}, \citenamefont {Van~de Graaff}, \citenamefont
  {Popowski}, \citenamefont {D'Incao}, \citenamefont {Julienne}, \citenamefont
  {Ye},\ and\ \citenamefont {Cornell}}]{Chapurin19}%
  \BibitemOpen
  \bibfield  {author} {\bibinfo {author} {\bibfnamefont {R.}~\bibnamefont
  {Chapurin}}, \bibinfo {author} {\bibfnamefont {X.}~\bibnamefont {Xie}},
  \bibinfo {author} {\bibfnamefont {M.~J.}\ \bibnamefont {Van~de Graaff}},
  \bibinfo {author} {\bibfnamefont {J.~S.}\ \bibnamefont {Popowski}}, \bibinfo
  {author} {\bibfnamefont {J.~P.}\ \bibnamefont {D'Incao}}, \bibinfo {author}
  {\bibfnamefont {P.~S.}\ \bibnamefont {Julienne}}, \bibinfo {author}
  {\bibfnamefont {J.}~\bibnamefont {Ye}}, \ and\ \bibinfo {author}
  {\bibfnamefont {E.~A.}\ \bibnamefont {Cornell}},\ }\href {\doibase
  10.1103/PhysRevLett.123.233402} {\bibfield  {journal} {\bibinfo  {journal}
  {Phys. Rev. Lett.}\ }\textbf {\bibinfo {volume} {123}},\ \bibinfo {pages}
  {233402} (\bibinfo {year} {2019})}\BibitemShut {NoStop}%
\bibitem [{\citenamefont {Massignan}\ and\ \citenamefont
  {Castin}(2006)}]{Yvan}%
  \BibitemOpen
  \bibfield  {author} {\bibinfo {author} {\bibfnamefont {P.}~\bibnamefont
  {Massignan}}\ and\ \bibinfo {author} {\bibfnamefont {Y.}~\bibnamefont
  {Castin}},\ }\href {\doibase 10.1103/PhysRevA.74.013616} {\bibfield
  {journal} {\bibinfo  {journal} {Phys. Rev. A}\ }\textbf {\bibinfo {volume}
  {74}},\ \bibinfo {pages} {013616} (\bibinfo {year} {2006})}\BibitemShut
  {NoStop}%
\bibitem [{\citenamefont {Zhang}\ and\ \citenamefont {Zhang}(2018)}]{Ren18}%
  \BibitemOpen
  \bibfield  {author} {\bibinfo {author} {\bibfnamefont {R.}~\bibnamefont
  {Zhang}}\ and\ \bibinfo {author} {\bibfnamefont {P.}~\bibnamefont {Zhang}},\
  }\href {\doibase 10.1103/PhysRevA.98.043627} {\bibfield  {journal} {\bibinfo
  {journal} {Phys. Rev. A}\ }\textbf {\bibinfo {volume} {98}},\ \bibinfo
  {pages} {043627} (\bibinfo {year} {2018})}\BibitemShut {NoStop}%
\bibitem [{\citenamefont {Zhang}\ and\ \citenamefont {Zhang}(2019)}]{Ren19}%
  \BibitemOpen
  \bibfield  {author} {\bibinfo {author} {\bibfnamefont {R.}~\bibnamefont
  {Zhang}}\ and\ \bibinfo {author} {\bibfnamefont {P.}~\bibnamefont {Zhang}},\
  }\href {\doibase 10.1103/PhysRevA.100.063607} {\bibfield  {journal} {\bibinfo
   {journal} {Phys. Rev. A}\ }\textbf {\bibinfo {volume} {100}},\ \bibinfo
  {pages} {063607} (\bibinfo {year} {2019})}\BibitemShut {NoStop}%
\bibitem [{\citenamefont {Xiao}\ \emph {et~al.}(2019)\citenamefont {Xiao},
  \citenamefont {Zhang},\ and\ \citenamefont {Zhang}}]{Xiao2019}%
  \BibitemOpen
  \bibfield  {author} {\bibinfo {author} {\bibfnamefont {D.}~\bibnamefont
  {Xiao}}, \bibinfo {author} {\bibfnamefont {R.}~\bibnamefont {Zhang}}, \ and\
  \bibinfo {author} {\bibfnamefont {P.}~\bibnamefont {Zhang}},\ }\href
  {\doibase 10.1007/s00601-019-1530-z} {\bibfield  {journal} {\bibinfo
  {journal} {Few-Body Systems}\ }\textbf {\bibinfo {volume} {60}},\ \bibinfo
  {pages} {63} (\bibinfo {year} {2019})}\BibitemShut {NoStop}%
\bibitem [{\citenamefont {Zhang}\ and\ \citenamefont {Zhang}(2020)}]{Ren20}%
  \BibitemOpen
  \bibfield  {author} {\bibinfo {author} {\bibfnamefont {R.}~\bibnamefont
  {Zhang}}\ and\ \bibinfo {author} {\bibfnamefont {P.}~\bibnamefont {Zhang}},\
  }\href {\doibase 10.1103/PhysRevA.101.013636} {\bibfield  {journal} {\bibinfo
   {journal} {Phys. Rev. A}\ }\textbf {\bibinfo {volume} {101}},\ \bibinfo
  {pages} {013636} (\bibinfo {year} {2020})}\BibitemShut {NoStop}%
\bibitem [{\citenamefont {Verhaar}\ \emph {et~al.}(1984)\citenamefont
  {Verhaar}, \citenamefont {van~den Eijnde}, \citenamefont {Voermans},\ and\
  \citenamefont {Schaffrath}}]{2dbpc}%
  \BibitemOpen
  \bibfield  {author} {\bibinfo {author} {\bibfnamefont {B.~J.}\ \bibnamefont
  {Verhaar}}, \bibinfo {author} {\bibfnamefont {J.~P. H.~W.}\ \bibnamefont
  {van~den Eijnde}}, \bibinfo {author} {\bibfnamefont {M.~A.~J.}\ \bibnamefont
  {Voermans}}, \ and\ \bibinfo {author} {\bibfnamefont {M.~M.~J.}\ \bibnamefont
  {Schaffrath}},\ }\href {\doibase 10.1088/0305-4470/17/3/020} {\bibfield
  {journal} {\bibinfo  {journal} {Journal of Physics A: Mathematical and
  General}\ }\textbf {\bibinfo {volume} {17}},\ \bibinfo {pages} {595}
  (\bibinfo {year} {1984})}\BibitemShut {NoStop}%
\bibitem [{aa2()}]{aa2d}%
  \BibitemOpen
  \href@noop {} {\emph {\bibinfo {title} {\rm Notice that the definitions of
  the 2D scattering length in Ref.~\cite{busch} and in this work are
  different.}}}\BibitemShut {Stop}%
\bibitem [{\citenamefont {Nishida}\ and\ \citenamefont {Tan}(2008)}]{shina}%
  \BibitemOpen
  \bibfield  {author} {\bibinfo {author} {\bibfnamefont {Y.}~\bibnamefont
  {Nishida}}\ and\ \bibinfo {author} {\bibfnamefont {S.}~\bibnamefont {Tan}},\
  }\href {\doibase 10.1103/PhysRevLett.101.170401} {\bibfield  {journal}
  {\bibinfo  {journal} {Phys. Rev. Lett.}\ }\textbf {\bibinfo {volume} {101}},\
  \bibinfo {pages} {170401} (\bibinfo {year} {2008})}\BibitemShut {NoStop}%
\bibitem [{\citenamefont {Bougas}\ \emph {et~al.}()\citenamefont {Bougas},
  \citenamefont {Mistakidis}, \citenamefont {Alshalan},\ and\ \citenamefont
  {Schmelcher}}]{bougas2020stationary}%
  \BibitemOpen
  \bibfield  {author} {\bibinfo {author} {\bibfnamefont {G.}~\bibnamefont
  {Bougas}}, \bibinfo {author} {\bibfnamefont {S.~I.}\ \bibnamefont
  {Mistakidis}}, \bibinfo {author} {\bibfnamefont {G.~M.}\ \bibnamefont
  {Alshalan}}, \ and\ \bibinfo {author} {\bibfnamefont {P.}~\bibnamefont
  {Schmelcher}},\ }\href@noop {} {}\Eprint {http://arxiv.org/abs/2001.10722}
  {arXiv:2001.10722} \BibitemShut {NoStop}%
\end{thebibliography}
%

\end{document}